\documentclass[aps,prx,onecolumn,superscriptaddress,11pt,longbibliography]{revtex4-1}

\usepackage{times}
\usepackage{graphicx}
\usepackage{siunitx}
\usepackage{setspace}
\usepackage{xcolor}
\usepackage{float}
\usepackage{array,multirow}
\usepackage{makecell}
 \usepackage{amsmath}

\begin{document}

\title{Vortex transition and thermal mixing by pitching a perforated flexible panel} 

\author{Yicong Fu}
\thanks{To whom correspondence should be addressed. \\E-mail: yf357@cornell.edu. \\Address: B45 Riley Robb Hall, 111 Wing Drive, Ithaca, New York 14853, USA. \\Phone: 434-466-7151.}
\affiliation{Sibley School of Mechanical and Aerospace Engineering, Cornell University, Ithaca, New York 14853, USA}

\author{Zhengyang Liu}
\affiliation{Department of Biological and Environmental Engineering, Cornell University, Ithaca, New York 14853, USA}

\author{Samir Tandon}
\affiliation{Department of Biological and Environmental Engineering, Cornell University, Ithaca, New York 14853, USA}

\author{Jake Gelfand}
\affiliation{Department of Biological Sciences, Cornell University, Ithaca, New York 14853, USA}

\author{Sunghwan Jung}
\thanks{To whom correspondence should be addressed. \\E-mail: sj737@cornell.edu.}
\affiliation{Department of Biological and Environmental Engineering, Cornell University, Ithaca, New York 14853, USA}

\date{\today}

\begin{abstract}
The effective transport of heat and mass is crucial to both industrial applications and physiological processes. Recent research has evaluated the benefit of using flexible reeds for triggering the vortex induced vibration to enhance mixing, as opposed to traditional techniques like rigid blender or static meshes. Inspired by the soft, porous, and moving fish gill lamellae, we proposed a new concept of thermal dispenser that prescribes active pitching motion to the leading edge of an otherwise passively flapping perforated panel. Experimental measurements revealed drastic differences between the steady leaky flow wake behind a statically deflected perforated panel and the periodic shedding wakes with complex vortex structure transitions behind an actuated perforated panel with or without chord-wise flexibility. A semi-empirical simulation of the thermal convection and diffusion takes the experimentally obtained velocity as input and yields the temperature results. Vortex dynamics, Lagrangian coherent structures, and thermal mixing behaviors were analyzed and compared to elucidate the effects of kinematics, perforation, and flexibility on the wake mode transitions, lateral entrainment mixing, and overall heating. Our work provides a foundational understanding of the fluid-structure interactions of perforated bendable panels under active control which has not been described before in the intermediate Reynolds number range. It provides insights for developing an innovative bio-inspired heat or mass dispenser potentially suitable for subtle and small scale applications.  
\end{abstract}

\maketitle

\section{Introduction}
    The material and thermal transport is a ubiquitous phenomenon governed by the convection and diffusion with crucial applications in both human technologies and biological activities, such as wastewater management \cite{tong2016global, vikas2015coastal}, temperature regulation \cite{khalaj2017review, wu2006combined}, blood dialysis \cite{himmelfarb2020current}, respiration  \cite{wood2021osmorespiratory}, predation \cite{hughes2010predators, wisenden2000olfactory}, and chemical communications \cite{reddy2022olfactory, chen2000human, grieves2022olfactory}. Tremendous research efforts has been put into the development of effective fluid mixing from microfluidic channels \cite{aref2002development, aref2020stirring} to large turbulent eddies \cite{dimotakis2005turbulent, caulfield2021layering}. Within the laminar flow regime, a major obstacle for mixing is the boundary layer formed due to the viscous no-slip condition at the solid-fluid interface. 
    
    Passive mixers have been proposed and used in many situations. Rigid vortex generators including protrusions and dimples on the surface can lead to the formation of vortices and turbulence that encourages mixing between the bulk and the boundary layer \cite{zhang2021combined, luo2016thermal, promvonge2010enhanced, he2021film, wang2015experimental, tang2016new, han2024insight}. However, the rigid vortex generators are known to significantly increase the pressure losses due to higher surface friction and geometric anomalies, requiring more energy consumption to maintain the desired flow rate. In recent years, flexible vortex generators have received much attention as a potential replacement for the rigid counterparts. By tuning the size \cite{liao2023experimental}, flexural rigidity \cite{zhong2022experimental, liu2024heat, lee2017heat}, mass distribution \cite{jin2024enhancing}, attachment condition \cite{chen2024enhancement, park2020heat}, and array placement \cite{chen2020heat, ali2016heat, ding2024improvement}, the vortical flow downstream of the flexible reed can significantly enhance the thermal mixing and local Nusselt number ($Nu$) in the case of boundary heating and cooling. It has been reported that the passively mounted flexible reed generally fall in three kinematic categories in response to the hydrodynamic forces: static reconfiguration, vortex induced vibration (VIV), and lodging \cite{zhang2020fluid}. Static reconfiguration occurs when the hydrodynamic forces are relatively small and steady due to the absence or weak formation of vortices. They behave similar to rigid panels since material deformation is minor and motion is negligible. VIV occurs when the vortices periodically shed from the edge of the reed at a certain frequency that locks in with the vibrational frequency of the solid structure. This mode generates much stronger and coherent vortices that contribute to the overall mixing. Lodging happens when the hydrodynamic forces completely push the flexible reed against the solid wall so that the resulting shape acts like a stationary hump. The limitations of passive flexible vortex generators are obvious. Their efficacy necessitates the occurrence of VIV, which is strongly dependent on the flow condition, resonance of the solid structure, and mounting boundary conditions, restraining the versatility of their applications. The passive nature still drains mechanical energy from the flow and result in pressure loss.

    Active mixing, on the other hand, has a wider application potential thanks to the conscious manipulation of the fluid flow, despite the drawback of energy consumption. Many examples exist in nature, including fish gill respiration \cite{hughes1972morphometrics}, insect wing-beat-induced odor capture \cite{li2018balance, lou2024wing}, honey bee hive's active ventilation \cite{peters2019collective}, cilia pumping \cite{gilpin2020multiscale}, cytoplasmic streaming \cite{shelley2024flows}, etc. Many bio-inspired devices have been made to mimic the coordinated motions of rigid and flexible structures \cite{den2008artificial, shields2010biomimetic, wang2023wirelessly, zhao2024thermal, nguyen2021enhancement, xie2020novel}. Experiments and simulations have shown their promising capabilities of material transport and mixing. In the context of gradient-driven flow, the concept of dynamic filtration has been proposed as a means to maintain high filtration flux across semi-permeable membranes by preventing membrane fouling with increased shear stresses \cite{jaffrin2012hydrodynamic}. In this work, we take inspiration from fish gill's active pumping respiration and integrate key features into our engineering model. Fish gill filaments form a mesh-like structure by the primary lamellae and the perpendicularly standing secondary lamellae. They have large surface areas and rich capillary vessels for metabolic exchange between blood and surrounding water \cite{hughes1972morphometrics}. Depending on the mechanism of water intake (ram ventilation or operculum pumping), gill filaments are found to have different degrees of calcification to fine tune the rigidity such that hydrodynamic forces do not create non-respiratory shunting flows \cite{turko2020calcified, strother2013hydrodynamic}. Dissections of the fish gills also identified abductor and adductor muscles attached to the filaments that allow active control of lateral motions \cite{hughes1972morphometrics}. Though complex, the key features of fish gill filaments can be reduced to three parameters: porosity, flexibility, and motion. Thus, we incorporated all three into our proposed thermal dispenser: a pitching perforated panel with chord-wise flexibility. 

    The fluid structure interactions (FSI) of actuated hydrofoils have received extensive research in the context of vortex formation, propulsion, and fluid pumping. Fish inspired oscillatory and undulatory swimming was thoroughly summarized by Smits \cite{smits2019undulatory}. Investigations in the dimensionless range of fish locomotion have reported a 2-pair (2P) to 2-single (2S) wake structure transition as a function of the amplitude-defined Strouhal number ($St_A = Af/U_\infty$), dimensionless wavelength ($\lambda^* = \lambda/c$), and aspect ratios ($\mathit{AR}$) \cite{dewey2012relationship, moored2012hydrodynamic, buchholz2006evolution, dong2005wake}. Here, $A$ is the oscillation amplitude, $f$ is the oscillation frequency, $U_\infty$ is the characteristic incoming flow velocity, $\lambda$ is the solid deformation wavelength, and $c$ is the chord length. Schnipper \textit{et al.} \cite{schnipper2009vortex} adopted 2D assumption in a soap film experiment with pitching rigid foil at the leading edge and summarized a phase map for wake transitions from 2S von Karman wake to 8P wake as a function of the dimensionless amplitude ($A_D = A/d$) and foil-thickness($d$)-defined Strouhal number ($St_D = df/U_\infty$), which can be combined as the amplitude-defined Strouhal number $St_A$. As a consequence of the wake vortex structure change, mean velocity streaming, force generation, hydrodynamic power consumption, and hydrodynamic efficiency were analyzed, but few has considered the thermal mixing benefit of these vortex structures. Recent research has also started to examine more complicated conditions of the foils such as irregular planform shapes \cite{van2017impact}, trailing edge (TE) tapers \cite{yeh2017efficient}, surface roughness \cite{dean2010shark}, and leaky surfaces \cite{bayazit2014perforated, marzin2022flow, li2021aerodynamics, kakroo2024high}. Particularly, leaky foils with perforations or bristle structures have raised huge interest among biomimetic engineers due to their outstanding performances in certain fluid and kinematic conditions. Statically positioned perforated flexible panels in cross flow have been reported to reduce drag and flexural rigidity ($\mathit{EI}$), and thus stabilize at a more-deformed reconfiguration angle than non-perforated panels \cite{guttag2018aeroelastic, jin2020distinct}. Clap-and-fling flight of small insects with bristled wings also revealed the benefit of drag reduction without significantly sacrificing lift \cite{kasoju2021aerodynamic, jones2016bristles}. In this work, we incorporate perforation, material flexibility, and prescribed motion to the FSI. We aim to characterize the different vortex shedding mechanisms and thermal mixing effects, potentially opening a new avenue for bio-inspired heat and mass dispensing system design. 

\section{Experimental methods}
    \subsection{Apparatus fabrication}
    
    \begin{figure}[htbp]
    \centering
    \includegraphics[width=1\textwidth]{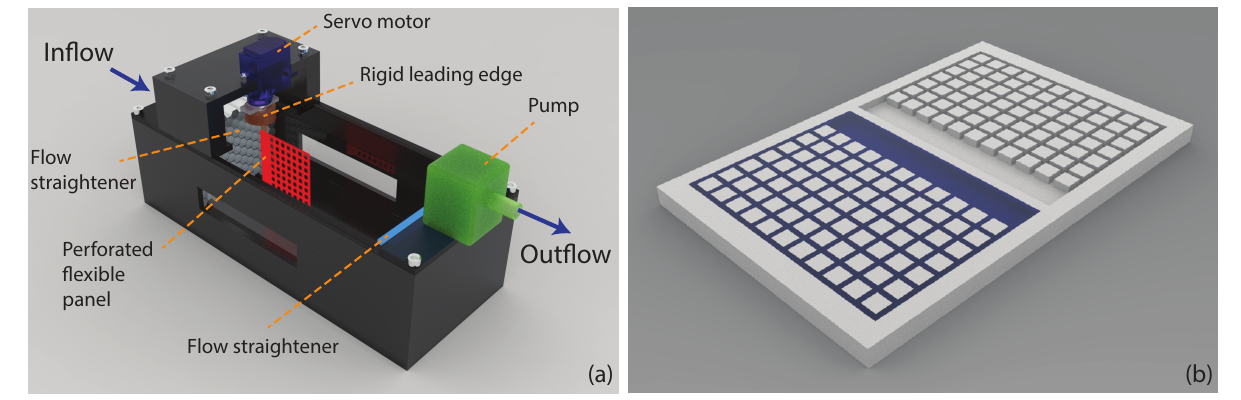}
    \caption{\textbf{Experimental apparatus rendered in Luxion Keyshot.} (a) In-house designed and fabricated flow channel. (b) Silicone mold of the perforated flexible panel. One side is filled with silicone (blue) and the other side is empty. } 
    \label{DIY flow tank render}
    \end{figure}

    We designed and fabricated a low-cost fluid channel (FIG. \ref{DIY flow tank render}(a)) capable of creating uniform, steady, and laminar background flows at different Reynolds numbers ($Re$). A Bambu Lab P1P 3D-printer was used to print polylactic acid (PLA) with FDM method. A rectangular channel of 186$\times$76$\times$54 mm (test section 90$\times$70$\times$50 mm; length$\times$width$\times$height) supports an overhanging mount for motors above the interface. A 30 mm thick honeycomb straightener with an average hole diameter of 5 mm was placed upstream to laminarize the inflow. A 3D-printed PLA straightener was placed downstream to further laminarize the fluid flow prior to the outflow sink. A 12 volt DC water pump was connected to the downstream straightener to generate flow. Two clear polycarbonate sheet (McMaster Carr) windows were cut open through the side walls of the channel to allow illumination. The channel was positioned in a larger fish tank with water filled to the upper edge of the channel. The water pump drew water from inside the channel and recirculated it in the fish tank. Outside the channel, plastic meshes were placed around the pump exit and the channel inlet as dampers to absorb the vibrations. An MG90S micro servo motor is mounted over the flow channel to actuate pitching motion about the center axis of the motor shaft. A 3D printed PLA rigid shaft extended vertically downward from the motor shaft to attach the perforated panels. 

    Two perforated panels with the same dimensions but different flexural rigidities ($EI$) were fabricated. The panel had an overall rectangular shape with a 45 mm span ($s$) and a 34.5 mm chord ($c$). Along the chord, 7 square perforation holes of $a\times a = 3\times3$ mm size were separated by $b = 1$ mm solid posts. Along the span, the chord-wise hole arrays were linearly repeated 11 times with $b = 1$ mm separations to result in a total of 77 perforation holes. The rigid panel was 3D printed with PLA. The flexible panel was casted with MoldStar 31T silicone rubber (SmoothOn Inc.) in a 3D printed PLA mold (FIG. \ref{DIY flow tank render}(b)). Both panels were 0.75 mm thick. From the same dimension, the second moment of area ($I$) is the same for both panels. The Young's modulus ($E$) is $2750\pm160$ MPa for PLA and approximately 2 MPa for silicone of A30 shore hardness. The $EI$ would have three orders of magnitude difference, hence the naming ``flexible'' and ``rigid''. After fabrication, the panel's non-perforated side is fused to the rigid shaft from the motor as the leading edge (LE). 
    
    \subsection{Particle image velocimetry}

    \begin{figure}[htbp]
    \centering
    \includegraphics[width=1\textwidth]{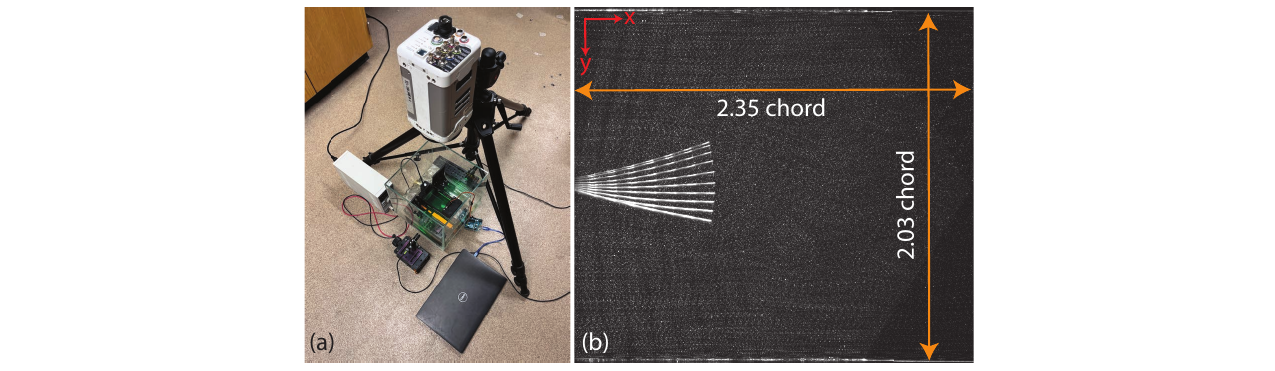}
    \caption{\textbf{Particle image velocimetry.} (a) Experimental setup showing the relative positions of the flow channel, high-speed camera, continuous laser sheets, and Arduino Uno microcontroller. (b) Overlaid every 10th image of one downstroke of the perforated flexible panel pitching at $f = 0.64$ Hz, showing the DPIV domain size. } 
    \label{PIV setup}
    \end{figure}

    2D digital particle image velocimetry (DPIV) was performed to capture the fluid velocity field in the experimental setup as FIG. \ref{PIV setup}(a). One Class IIIB laser (LaserLand) was placed on each side of the channel to create a shadowless horizontal sheet shining through the windows. A NOVA S6 Photron high-speed camera was positioned vertically downward above the test section to capture the fluid motion at 125 frames per second (FPS) through the PFV4 software. Channel was filled with DI water and seeded with Dantec silver-coated hollow glass spheres (10 $\mu$m). The recording domain is defined as 2.03 $c$ wide and 2.35 $c$ long, containing all the perforation holes and separation posts (FIG. \ref{PIV setup}(b)). 
    
\begin{figure}[htbp]
    \centering
    \includegraphics[width=1\textwidth]{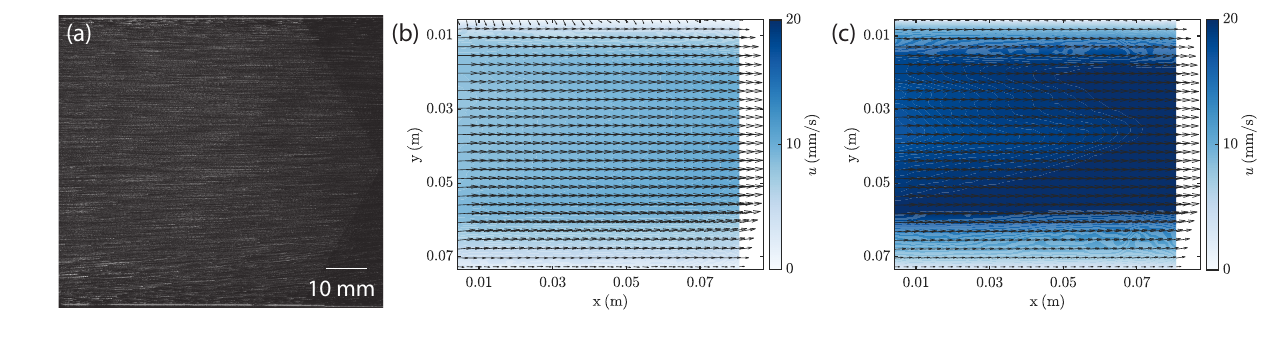}
    \caption{\textbf{Uniform and steady laminar flow.} (a) Laser-illuminated neutrally buoyant particle pathlines over 1 second. (b) Velocity field at 4 VDC Pump speed. (c) Velocity field at 12 VDC Pump speed.} 
    \label{Uniform field}
    \end{figure}

    Without putting anything in the test section, we confirmed the background flow to be uniform, steady, and laminar. An overlay of the DPIV images (FIG. \ref{Uniform field}(a)) reveal the pathlines of the seeding particles to follow the laminar streaks from left to right. DPIV results from two extreme flow speed cases (FIG. \ref{Uniform field}(b-c)) show consistently reliable flow fields with a wide uniform flow profile with relatively small boundary layers near the wall.
    
    During experiments with the panel, the laser sheet's height was adjusted to align with the perforation holes at the center of the span. This helps visualize all the fluid flow penetrating through the perforation and guarantees negligible 3D effects from the top and bottom edges. The pump was set to 12 VDC, which creates an averaged flow speed of 17 mm/s equivalent to a fixed $Re = \rho U_\infty c/\mu$ of 590. Either panel was pitched about the leading edge at a frequency $f$ ranging from 0.64 to 1.14 Hz. For every condition, a minimum 2 min wait time was given before starting recording to guarantee the flow stabilized to a repeating pattern without effects from the previous condition or external sources. For steady flow fields without solid structures or with a static structure, 8 s (1000 frames) was recorded. For unsteady flow fields around the pitching panel, 10 stroke cycles were recorded to extract the phase-averaged velocity to multiple frames. Image postprocessing were done in ImageJ, including outlier removal, de-speckling, contrast enhancement, background subtraction, and binarization, to generate the masks for the solid structures in the DPIV video. DPIV analysis were performed in PIVlab version 2.63 \cite{stamhuis2014pivlab} in MATLAB 2024b on the raw video, with masks overlaid. With a 2-layer interrogation window and a 50\% stepping, we achieved $10\times10$ px velocity field resolution from the $1024\times1024$ px image resolution. 
    
\section{Computational methods}
    \subsection{Convection Diffusion Equation}
    Due to the difficulty of experimentally fabricating a thermal couple or a heat exchanger of this shape and flexibility, we decided to adopt a semi-empirical computation method to simulate the thermal mixing by using the experimental flow field as an input. Conventionally, full computational fluid dynamics (CFD) methods for the present scenario would require the integration of three governing physics: fluid motion, solid deformation, and heat transfer. The first two are usually coupled as an iterative feedback loop since the fluid flow imposes hydrodynamic forces to the solid structure, and the deforming solid structure in turn dictates the motion of the fluid. Such computational FSI has been well established to study flexible vortex generators \cite{shoele2014computational, kakroo2024high, chen2020heat}. However, the FSI is usually computationally decoupled from the heat transfer simulations which is based on the Convection Diffusion equation
    
    \begin{equation}\label{eq:ConvDiff}
    \frac{\partial T}{\partial t} = D\boldsymbol{\nabla}^2T-\boldsymbol{u}\cdot\boldsymbol{\nabla} T \, .
    \end{equation}
    
    \noindent Here, $T$ is the fluid temperature, $t$ is time, $\boldsymbol{u} = [u,v]$ is the 2D fluid velocity vector, and $D$ is the thermal diffusivity of water taken as $1.43\times10^{-7}~\mathrm{m}^2/\mathrm{s}$. Since the DPIV experiments essentially provide the same $\boldsymbol{u}$ information as the direct numerical solutions from an FSI solver, we believe that this semi-empirical method is able to yield meaningful and accurate results on thermal mixing. 
    
    \subsection{Simulation settings and validations}
    The preparation for the thermal simulation starts from pre-processing the experimental velocity field and solid structure positions. For steady flow fields with static solid structures in water, no velocity averaging was performed. The raw velocity fields across 1000 frames were repeated 4 times to extend the simulation time to 32 seconds, which is enough to reach a steady-state equilibrium solution. For unsteady flow fields around the pitching panel, phase-averaged velocity fields were extracted from 10 recorded stroke cycles and repeated $N$ times such that the total simulation time also extends to about 32 seconds. Here, $N$ is equivalent to the number of phase-averaged stroke cycles simulated. Similarly, the solid structure position for a static object is extracted from images, mapped to the same meshgrid as the velocity field, and kept unchanged throughout the simulation. The solid structure positions of the moving and deforming panels were extracted from the phase-averaged frames, mapped to the velocity meshgrids, and repeated $N$ times. 
    
    In this work, we assume that the solid structure immersed in the fluid is not only a vortex generator but also the heat source responsible for increasing the fluid temperature. We artificially impose a Dirichlet constant $T$ boundary condition across the background flow inlet (left) at $T_\infty = 300$ K at all times. This intuitively simulates the condition of a one-directional flow through a pipeline section at a constant source temperature. We also impose another Dirichlet constant $T$ boundary condition at the solid structure heat sources at $T_\mathrm{s} = 350$ K. Thus, the maximum temperature difference is $\Delta T = 50$ K. At the top, bottom, and outlet (right) boundaries, we impose Neumann constant flux boundary conditions, $\partial T/\partial \boldsymbol{n}$ = 0, with $\boldsymbol{n}$ being the outward normal vector at each boundary surface. The interior nodes are computed based on the temperature and velocity information of their adjacent nodes. Specifically, $\boldsymbol{\nabla} T$ of the interior nodes follow the first order upwind scheme, as

    \begin{subequations} \label{eq:upwind}
    \begin{align}
    \left( \frac{\partial T}{\partial x} \right)_{i,j} &\approx
    \begin{cases}
    \frac{T_{i,j} - T_{i-1,j}}{\Delta x}, & \text{if } u_{i,j} > 0 \quad  \\
    \frac{T_{i+1,j} - T_{i,j}}{\Delta x}, & \text{if } u_{i,j} < 0 \quad 
    \end{cases} \\
    \left( \frac{\partial T}{\partial y} \right)_{i,j} &\approx
    \begin{cases}
    \frac{T_{i,j} - T_{i,j-1}}{\Delta y}, & \text{if } v_{i,j} > 0 \quad  \\
    \frac{T_{i,j+1} - T_{i,j}}{\Delta y}, & \text{if } v_{i,j} < 0 \quad 
    \end{cases}
    \end{align}
    \end{subequations} \, ,

    \noindent to prevent nonphysical oscillations in the computational results. $\boldsymbol{\nabla}^2 T$ of the interior nodes follow the central difference scheme, as 

    \begin{subequations} \label{eq:centralDiff}
    \begin{align}
    \left( \frac{\partial^2 T}{\partial x^2} \right)_{i,j} &\approx \frac{T_{i+1,j} - 2T_{i,j} + T_{i-1,j}}{\Delta x^2}\\
    \left( \frac{\partial^2 T}{\partial y^2} \right)_{i,j} &\approx \frac{T_{i,j+1} - 2T_{i,j} + T_{i,j-1}}{\Delta y^2}
    \end{align}
    \end{subequations} \, .

    \noindent Then, the remaining temperature gradient and Laplacian on the boundary nodes are simply extrapolated from the two adjacent interior nodes. The constructed $\partial T/\partial t$ matrix is then linearized to be integrated by MATLAB function ODE45 based on the explicit Runge-Kutta (4,5) formula \cite{shampine1997matlab, dormand1980family}. Simulation time domain is segmented based on the phase-averaged DPIV frames, making each segment 8 ms. Within each time segment, the integration time step sizes are automatically determined by ODE45, and only the temperature field at the end of the time segment is recorded as the result for the corresponding phase-averaged DPIV frame. Velocity fields within the time segment are linearly interpolated using the bounding fields. For the first time segment, the initial condition is set to $T_\mathrm{s} = 350$ K for all the solid structure nodes and $T = 300$ K for all the fluid nodes. For the following time segments, the previous recorded temperature field is then used as the initial condition for the next time segment to evolve the thermal mixing over time. 
    
    \begin{figure}[htbp]
    \centering
    \includegraphics[width=1\textwidth]{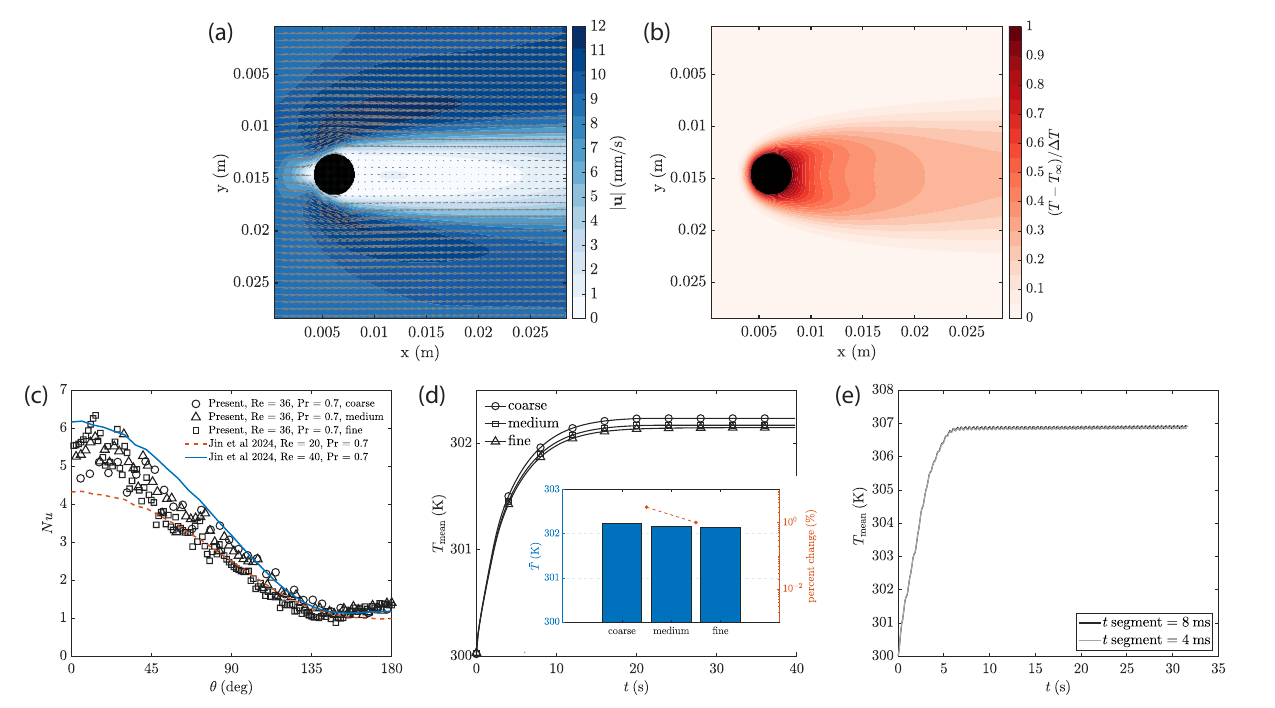}
    \caption{\textbf{Solver validation results.} (a) time-averaged velocity field of the flow past cylinder at $Re = 36$. (b) Simulated temperature field for the flow past cylinder at $Re = 36$. (c) Nusselt number along the solid-fluid interface at the cylinder. Upper and lower semi-circles are averaged due to the symmetry. Our results are compared against published results by Jin \textit{et al}. \cite{jin2024enhancing}, with permission from the corresponding author. (d) Domain-averaged temperature evolution over time for different meshgrid sizes in flow past cylinder. Inset compares the equilibrium temperatures (blue) for three meshgrid sizes and the percent changes (orange) to confirm acceptable mesh size independence beyond ``medium''. (e) Simulation time step size ($t$ segment length) convergence verified with the flexible panel pitching at $f = 1.14$ Hz.} 
    \label{solver validation}
    \end{figure}

    To validate the Convection Diffusion solver, we performed a canonical example of a forced convection by an isothermal circular cylinder in cross flow. We placed a 3D printed PLA circular cylinder with a 4 mm diameter ($d$) vertically in the middle of the test section with a constant $U_\infty$ about 9 mm/s. Taking water density as $\rho = 1000\: \mathrm{kg}/\mathrm{m}^3$ and dynamic viscosity as $\mu = 1\times10^{-3}\: \mathrm{Pa}\cdot\mathrm{s}$, the $Re = \rho U_\infty d/\mu$ is about 36. According to the dimensionless form of the Convection Diffusion equation, 

    \begin{equation}\label{eq:ConvDiff_nondim}
    \frac{\partial T^*}{\partial t^*} = \frac{1}{Pe}\boldsymbol{\nabla}^{*2}T^*-\boldsymbol{u}^*\cdot\boldsymbol{\nabla}^* T^* \, ,
    \end{equation}
    
    \noindent where $Pe$ denotes the Peclet number and the $(\cdot)^*$ denotes the non-dimensionalization, we understand that both $Re$ and the Prandtl number ($Pr$) characterize the physics since $Pe = RePr$. For a good comparison with published results, similar dimensionless numbers are crucial for maintaining the dynamic similarity. Thus, we artificially set the thermal diffusivity to $1.43\times10^{-6}\:\mathrm{m^2}/\mathrm{s}$ (one order higher than that of water) to force the $Pr$ at 0.7. [This setting is only for validating the solver. The true thermal diffusivity of water is used in other simulations of the present work.] Due to the horizontal 2D assumption, we neglect buoyancy effects from the fluid density change with temperature and set the Grashof number ($Gr$) and Richardson number ($Ri$) to 0. This will be true for simulations for the pitching panels too.
    
    As shown in FIG. \ref{solver validation}(a), the steady recirculating twin vortex wake behind the cylinder at $Re$ of 36 is consistent with previous reports \cite{taneda1956experimental}. Same simulation settings were applied, besides the isothermal condition on the cylinder is kept at $T_\mathrm{s} = 320$ K ($\Delta T = 20$ K). After reaching the steady-state solution, we examine the Nusselt number ($Nu$) along the semi-circle solid-liquid interface as an indicator for the ratio between heat convection and conduction, commonly used to validate simulations \cite{jin2024enhancing, park2017simulation, dennis1968steady}. Based on the following equation \cite{jin2024enhancing}, 

    \begin{equation}\label{eq:Nusselt}
    Nu|_\theta = -\frac{D}{\Delta T}\frac{\partial T}{\partial r}\Bigg|_\theta \, ,
    \end{equation}
    
    \noindent we compute the $Nu$ as a function of $\theta$ which is defined as 0$^\circ$ at the upstream stagnation center and positive in the clockwise direction. $\Delta T = 20$ K is the maximum temperature difference and $\partial T/\partial r$ is the radial temperature gradient along the solid surface at various $\theta$ angles. Our results in FIG. \ref{solver validation}(b-c) show excellent agreement with published data, exhibiting the same magnitude and trend. For smaller $\theta$, i.e. left half of the cylinder, $Nu$ values are more scattered due to the compression of the velocity boundary layer and the incapability of DPIV to experimentally resolve thin velocity boundary layers or reflect the no-slip condition. As a result, the thermal boundary layer on the upstream side is also compressed to be thinner, leaving a sharper temperature gradient potentially causing the data scatter. Acknowledging the limitation of the experimental data, we still believe the $Nu$ trend validates our solver's ability to study thermal mixing. 

    We also checked the meshgrid size independence threshold to reduce the computational cost while yielding converged results. The baseline mesh comes from the discretizations of the DPIV velocity field, which is inherently a Cartesian system. The baseline mesh size is about $d/12$. Linearly interpolating once, twice, and three times would result in smaller mesh sizes of $d/24$ (coarse), $d/48$ (medium), and $d/96$ (fine). All three mesh densities converged to highly similar results for the $Nu$ and domain-averaged temperature ($T_\mathrm{mean}$), as shown in FIG. \ref{solver validation}(c-d). Here, $T_\mathrm{mean}$ is calculated as $\frac{1}{S}\int T dx dy$ where $S$ is the area of the domain. As shown in the FIG. \ref{solver validation}(d) inset, the percent change of the equilibrium temperature ($\bar{T}$) from coarse to medium mesh is 3\% and from medium to fine mesh is 1\%. Thus, we decided to use the medium mesh density (twice interpolation) for all further simulations. Here, $\bar{T}$ is calculated as $\frac{1}{\Delta t}\int T_\mathrm{mean} dt$, where $\Delta t$ is the time length for time-averaging. 

    We further checked the simulation time step size convergence. The way we setup the pitching panel simulation assumes that the solid structure heats up the fluid replenishing its previous position to the maximum temperature when it moves to the next position. This setting relies on the premise of large overlap of solid structure positions between adjacent frames. To confirm the recording FPS is sufficiently high and the solid structure displacement between adjacent simulation frames is sufficiently small, we tested two time segment lengths with the flexible panel pitching at $f = 1.14$ Hz. The aforementioned 8-ms time segment is halved to 4 ms. Simulation results from both matched very well, with $\bar{T}$ reaching 306.894 K and 306.877 K respectively (FIG. \ref{solver validation}(e)). This test case is at the highest frequency with the least anticipated solid structure overlap between adjacent frames. Therefore, we safely choose 8 ms as the time segment for further simulations. 
    
\section{Results}
    \subsection{Static deflection at maximum pitch angle of 15$^\circ$}
    
    \begin{figure}[htbp]
    \centering
    \includegraphics[width=1\textwidth]{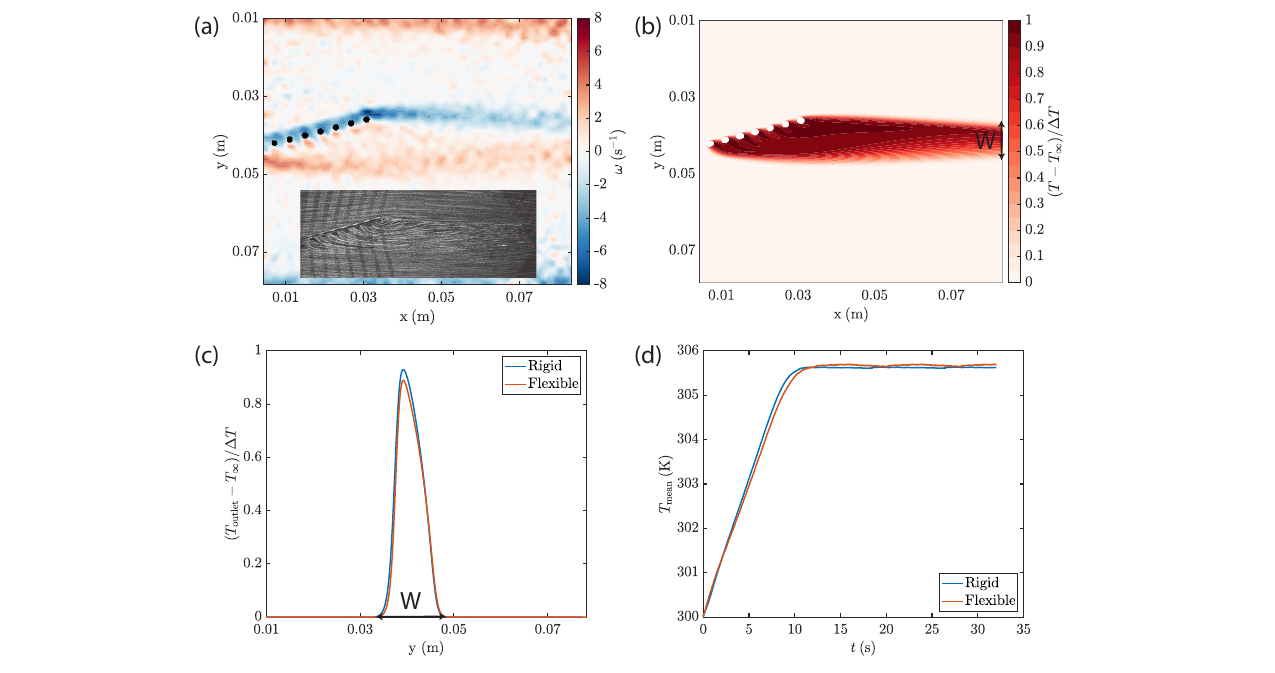}
    \caption{\textbf{Statically positioned rigid and flexible panel at 15$^\circ$ show nearly identical results.} (a) Instantaneous vorticity field around the flexible panel. (b) Time-averaged equilibrium temperature field around the flexible panel. (c) Outlet temperature wake widths for both panels. (d) Simulated domain-averaged temperature evolution over time for both panels. } 
    \label{static 15deg}
    \end{figure}

    As a reference, we first position the panel at the maximum pitching angle (15$^\circ$) without prescribing motion, thus naming it the ``static'' case. There is negligible difference between the rigid and flexible panels under this condition. The flexible panel experiences a very small hydrodynamic force due to the perforation letting water pass through, resulting in undetectable deformation compared with the rigid panel. No structural vibrations or periodic vortex shedding were seen, indicating that this belongs to the ``lodging'' regime, as reported in previous work \cite{zhang2020fluid}. For this steady flow field, we visualize the instantaneous vorticity and the formation of a low-velocity zone surrounded by shear layers (FIG. \ref{static 15deg}(a)). The time-averaged temperature field over the last 4 s of the simulation displayed little mixing in and outside the wake region due to insulation from the shear layers (FIG. \ref{static 15deg}(b)). The width of outlet thermal wake ($W$) was found to be nearly identical for both panels (FIG. \ref{static 15deg}(c)). Finally, the time evolution of $T_{\mathrm{mean}}$ reached $\bar{T}$ at 305.7 K at around 11 s for both panels (FIG. \ref{static 15deg}(d)). Consistent thermal performances between two panels demonstrate the same fluid flow and mixing ability at this particular setting. As a baseline reference, their similarity guarantees the differences among later cases with prescribed motion are due to the different unsteady FSI of two panels. These results would help quantify how much performance boost takes place when the panels are actuated. 
    
    \subsection{Kinematics of the pitching panels}

    \begin{figure}[htbp]
    \centering
    \includegraphics[width=1\textwidth]{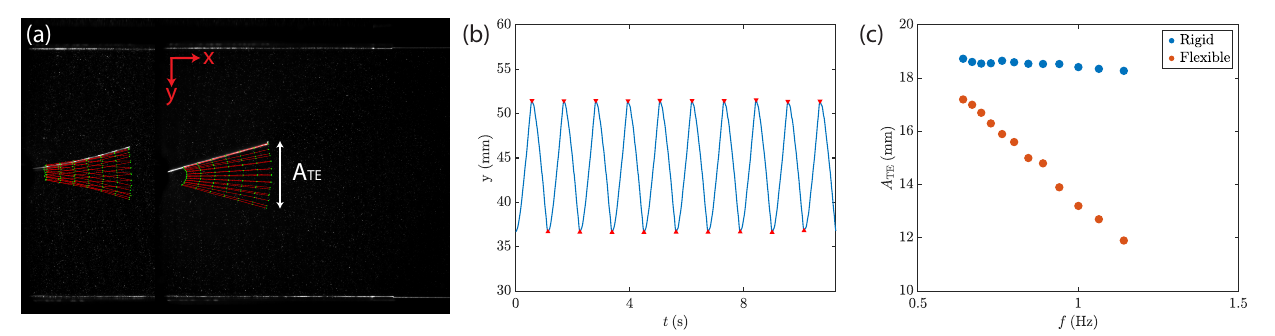}
    \caption{\textbf{Dynamic motions extracted by tracking.} (a) Centerline (red) of the panel top view with solid post locations (green) marked on the flexible (left) and rigid (right) panel pitching at $f = 0.89$ Hz. (b) Example of TE motion tracking from the flexible panel pitching at $f = 0.89$ Hz. (c) Compiled $A_\mathrm{TE}$ as a function of $f$ for both panels. } 
    \label{A_TE}
    \end{figure}

    Hereafter, we consider the pitching panels. We first characterized the kinematics of the rigid and flexible panels by tracking their motion in the 2D top view. As shown in FIG. \ref{A_TE}(a), the centerline of the panel (Red) was traced out from the DPIV masks and the locations of the solid posts (Green) separating the perforation holes were evenly distributed along the centerline according to the actual dimensions. The rigid panel sweeps out a true circular sector with negligible chord-wise deformation, while the flexible panel sweeps over a smaller funnel-shaped area due to bending. The TE positions were tracked using MATLAB package DLTdv8 \cite{hedrick2008software} and the subtraction between peaks and troughs defines the peak-to-peak amplitude ($A_{\mathrm{TE}}$) (FIG. \ref{A_TE}(b)). In summary, the rigid panel experiences little deformation across the tested $f$ range, whereas the flexible panel undergoes more bending and smaller $A_{\mathrm{TE}}$ at higher $f$ (FIG. \ref{A_TE}(c)). For reference, kinematic conditions and $St_A$ are summarized in Table \ref{Table St}. 

    \setlength{\tabcolsep}{10pt}
    \begin{table}[h]
    \centering
    \begin{tabular}{|c|c|c|c|c|}
        \hline
        $f$ (Hz) & Panel & $A_\mathrm{TE}$ (mm) & $U_\infty$ (mm/s) & $St_A$ \\ 
        \hline
        0.64  & Rigid  & 18.7 & 17.0 & 0.647 \\ 
        \cline{2-3}\cline{5-5}
               & Flexible  & 17.2 &  & 0.706 \\ 
        \hline
        0.67  & Rigid  & 18.6 & 17.0 & 0.670 \\ 
        \cline{2-3}\cline{5-5}
               & Flexible  & 17.0 &  & 0.733 \\ 
        \hline
        0.70  & Rigid  & 18.5 & 17.0 & 0.684 \\ 
        \cline{2-3}\cline{5-5}
               & Flexible  & 16.7 &  & 0.761 \\ 
        \hline
        0.73  & Rigid  & 18.6 & 17.0 & 0.699 \\ 
        \cline{2-3}\cline{5-5}
               & Flexible  & 16.3 &  & 0.794 \\ 
        \hline
        0.76  & Rigid  & 18.7 & 17.0 & 0.713 \\ 
        \cline{2-3}\cline{5-5}
               & Flexible  & 15.9 &  & 0.837 \\ 
        \hline
        0.80  & Rigid  & 18.6 & 17.0 & 0.732 \\ 
        \cline{2-3}\cline{5-5}
               & Flexible  & 15.6 &  & 0.876 \\ 
        \hline
        0.84  & Rigid  & 18.5 & 17.0 & 0.745 \\ 
        \cline{2-3}\cline{5-5}
               & Flexible  & 15.0 &  & 0.921 \\ 
        \hline
        0.89  & Rigid  & 18.5 & 17.0 & 0.775 \\ 
        \cline{2-3}\cline{5-5}
               & Flexible  & 14.8 &  & 0.971 \\ 
        \hline
        0.94  & Rigid  & 18.5 & 17.0 & 0.769 \\ 
        \cline{2-3}\cline{5-5}
               & Flexible  & 13.9 &  & 1.026 \\ 
        \hline
        1.00 & Rigid  & 18.4 & 17.0 & 0.780 \\ 
        \cline{2-3}\cline{5-5}
               & Flexible  & 13.2 &  & 1.084 \\ 
        \hline
        1.06 & Rigid  & 18.3 & 17.0 & 0.798 \\ 
        \cline{2-3}\cline{5-5}
               & Flexible  & 12.7 &  & 1.149 \\ 
        \hline
        1.14 & Rigid  & 18.3 & 17.0 & 0.797 \\ 
        \cline{2-3}\cline{5-5}
               & Flexible  & 11.9 &  & 1.228 \\ 
        \hline
    \end{tabular}
    \caption{Summary of the kinematics and $St_A$.}
    \label{Table St}
    \end{table}
    
    \subsection{Vortex structure transitions}

    Pitching panels are known to shed different vortex structures into the wake region depending on the oscillation frequency, amplitude, and the oncoming flow speed, namely $St_A$. The phase diagrams for wake structure transitions of pitching non-perforated rigid hydrofoils in 2D has been documented \cite{schnipper2009vortex}. However, little has been described for perforated flexible hydrofoils, hence our effort to examine the vortical structures and the transitions as a function of $f$ in our system. 

\begin{figure}[htbp]
    \centering
    \includegraphics[width=1\textwidth]{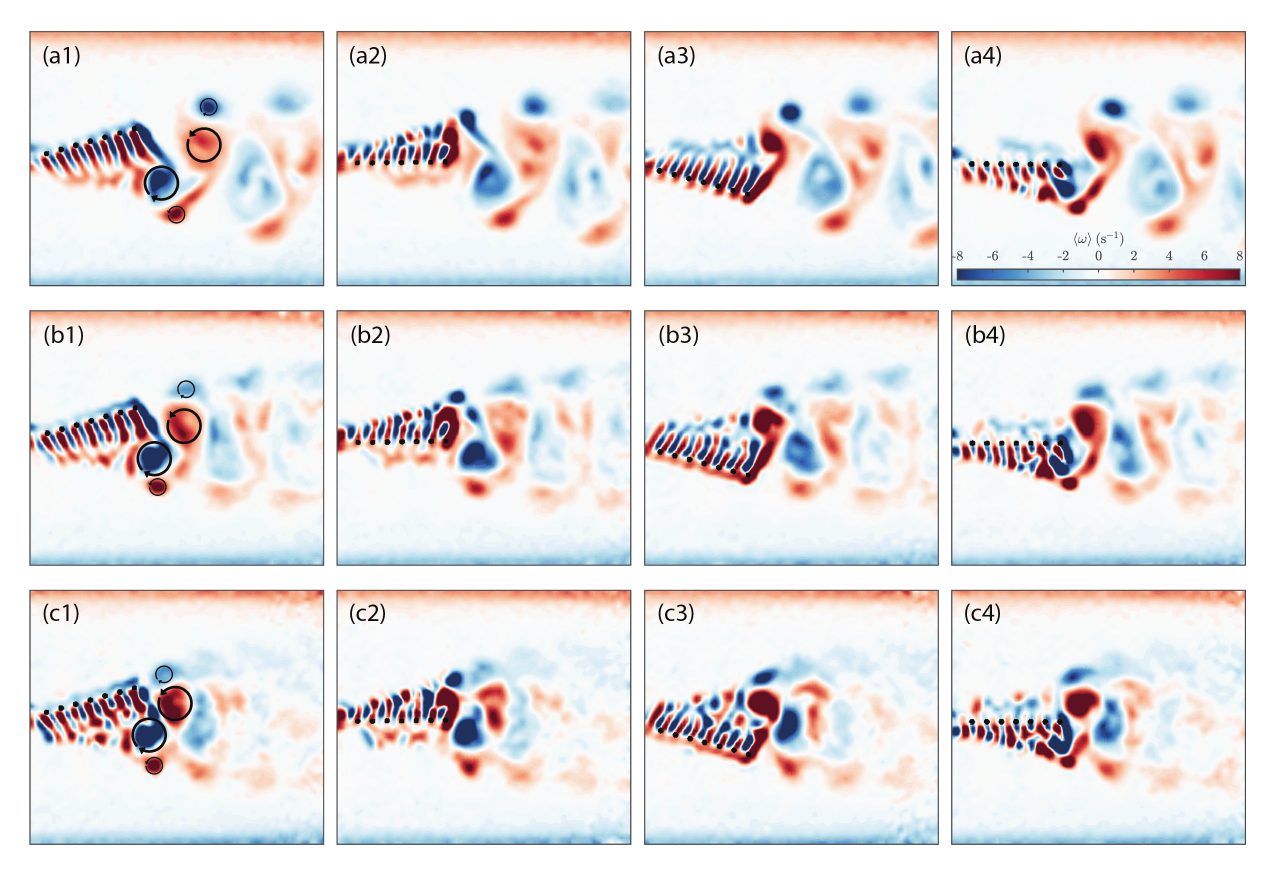}
    \caption{\textbf{Wake vortex structure transition illustrated by phase-averaged snapshots from DPIV with the rigid panel.} (a) A coherent 2P wake at $f = 0.64$ Hz, similar to the flexible panel at the same $f$. (b) A similar 2P wake at $f = 0.89$ Hz, with the vortex pair separation reduced. (c) A very localized 2P wake at $f = 1.14$ Hz, with very small vortex pair separation leading to continuous shear layers bounding the wake. (1-4) Numbers correspond to 0, 25, 50, 75\% progress of a phase-averaged cycle. \textbf{Note:} Red is counter-clockwise; Blue is clockwise. $\langle \omega \rangle$ denotes the phase-averaged vorticity.} 
    \label{vorticity rigid}
    \end{figure}

\begin{figure}[htbp]
    \centering
    \includegraphics[width=1\textwidth]{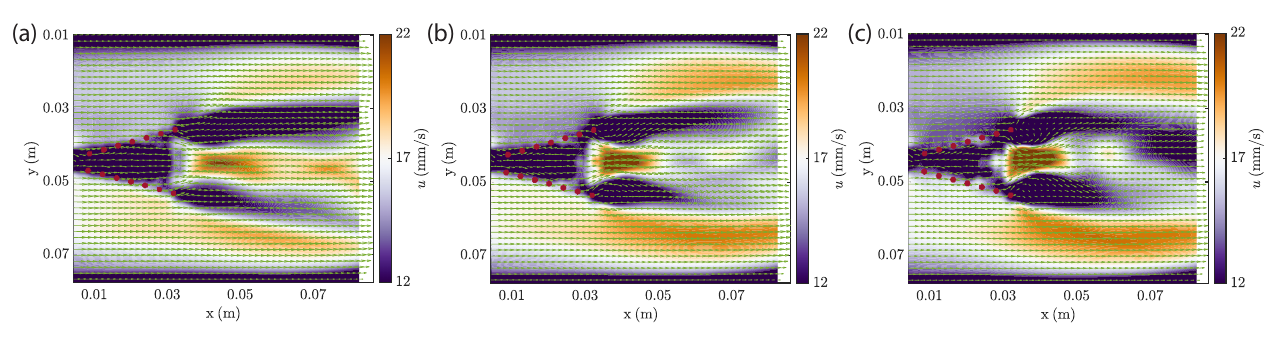}
    \caption{\textbf{Time-averaged velocity field showing the mean flow behavior contributing to the fluid convection in the wake of the rigid panel.} (a) $f = 0.64$ Hz, (b) $f = 0.89$ Hz, and (c) $f = 1.14$ Hz. Red dots outline the sweeping range.} 
    \label{mean velocity rigid}
    \end{figure}

    Across the tested $f$ range, the rigid panel did not exhibit a drastic change in the vortex shedding mode. A 2P wake was consistent throughout all cases, meaning that every pitching cycle generated 2 pairs of counter-rotating vortices (FIG. \ref{vorticity rigid}). Specifically, during each stroke, a large primary vortex and a small secondary vortex were created at the TE. The large primary vortex pairs with the small secondary vortex from the previous stroke. The smaller secondary vortex pairs with the large primary vortex from the next stroke. The relative position of the counter-rotating vortices in each pair always puts the large primary vortex closer to the center of the wake and the small secondary vortex on the outside. Each pair tends to squeeze a jet in the opposite direction to the oncoming background flow, leading to a reduced $x$ velocity, $u$, along the bounding regions of the wake (FIG. \ref{mean velocity rigid}). The secondary vortex is weaker in strength and smaller in size than the primary vortex, thus dissipating more quickly. At $f = 0.64$ Hz (FIG. \ref{vorticity rigid}(a)), the vortex pair has significant time to advect downstream before the creation of the next pair. This reduces the vortex dissipation and annihilation. As a result, vortices stay coherent for a long period of time and gets shed in the streamwise direction. The wake has no bounding shear layers on the side. The decent spatial separation between vortex pairs acts like an entrance that allows the fluid outside the wake to be entrained into the wake for potentially more thermal mixing. At $f = 0.89$ Hz (FIG. \ref{vorticity rigid}(b)), the same effects become less pronounced. The lateral fluid entrainment entrance becomes smaller due to the closer spacing between vortex pairs. The outer secondary vortex starts to evolve into a shear layer as it advects downstream. At $f = 1.14$ Hz (FIG. \ref{vorticity rigid}(c)), the outer secondary vortex quickly stretches and diffuses into a continuous shear layer due to the advection by the mean flow, rather than staying as a coherent structure. The formation of the bounding shear layers prohibits lateral entrainment and mixing. The inner primary vortex also disappears more quickly due to strong interactions and annihilation with the previous pitching cycle, resulting in little $u$ increment in the wake center except for a small region close to the TE (FIG. \ref{mean velocity rigid}(c)). 
    
\begin{figure}[htbp]
    \centering
    \includegraphics[width=1\textwidth]{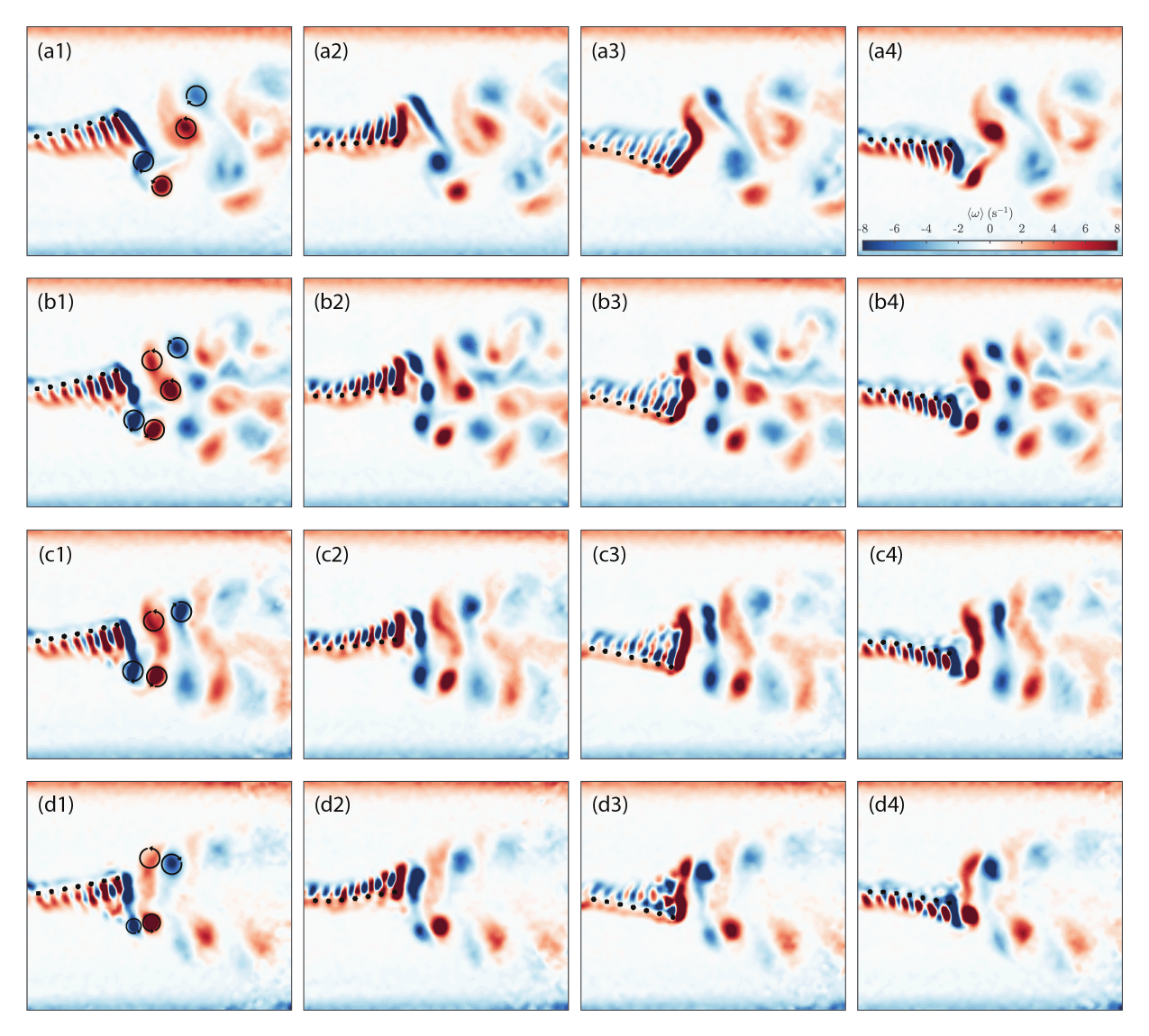}
    \caption{\textbf{Wake vortex structure transition illustrated by phase-averaged snapshots from DPIV with the flexible panel.} (a) A non-bifurcating 2P wake at $f = 0.64$ Hz with an orientation opposing the mean flow, similar to the rigid panel at the same $f$. (b) A 2P+2S wake at $f = 0.89$ Hz creating oscillating flow profiles across the width of the wake. (c) A bifurcating 2P wake at $f = 1.00$ Hz with the orientation transitioning from slightly opposing to slightly accelerating the mean flow. (d) A bifurcating 2P wake at $f = 1.14$ Hz with an orientation accelerating the mean flow. (1-4) Numbers correspond to 0, 25, 50, 75\% progress of a phase-averaged cycle. \textbf{Note:} Red is counter-clockwise; Blue is clockwise. $\langle \omega \rangle$ denotes the phase-averaged vorticity.} 
    \label{vorticity flexible}
    \end{figure}

\begin{figure}[htbp]
    \centering
    \includegraphics[width=1\textwidth]{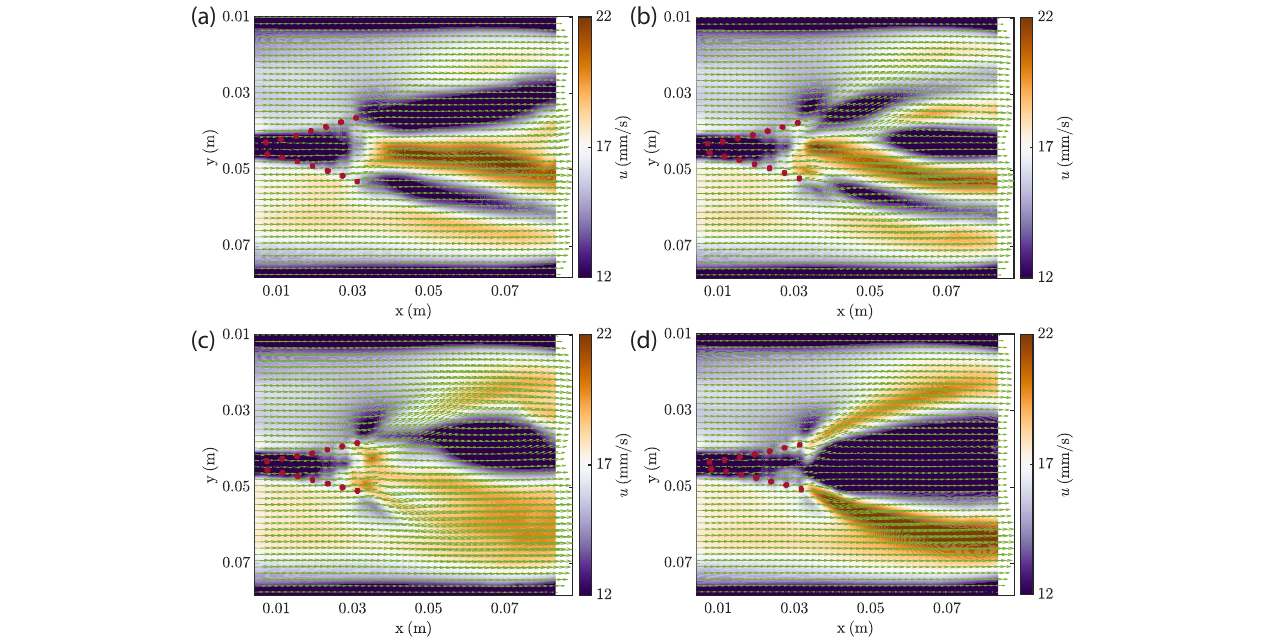}
    \caption{\textbf{Time-averaged velocity field showing the mean flow behavior contributing to the fluid convection in the wake of the flexible panel.} (a) $f = 0.64$ Hz, (b) $f = 0.89$ Hz, (c) $f = 1.00$ Hz, and (d) $f = 1.14$ Hz. Red dots outline the sweeping range.} 
    \label{mean velocity flexible}
    \end{figure}

    Conversely, pitching the flexible panel across the tested $f$ range had drastic changes in the vortex shedding mode. As $f$ increase from 0.64 Hz to 1.14 Hz, the vortex structure transitions from a non-bifurcating 2P wake to a 2P+2S wake and then to a bifurcating 2P wake. Due to the flexible material, the panel deflects into a more hydrodynamically favorable shape during the stroke reversal and allows water to flow more easily in the lateral direction at the TE. Another way to conceptually understand this difference is the Kutta condition: the flow cannot realistically wrap around a sharp TE and thus has to exit the TE smoothly. The flexible panel's bending at the stroke reversal increases the slope of the curvature near the TE, making it point more laterally than the rigid panel. This leads to a stronger lateral component of the fluid velocity vector and a wider wake region than that behind the rigid panel. At $f = 0.64$ Hz (FIG. \ref{vorticity flexible}(a)), a non-bifurcating 2P wake occurs. Every stroke, a long golf-club-shaped vortex is created at the TE. There is a strong concentrated vortex core first generated during the initial sweep of the stroke and followed by a slender vortex sheet that stays connected between the vortex core and the TE. They are shed together after the stroke recovery to be advected downstream while being stretched even more laterally. Eventually, they sever into two distinct vortices to contribute to different vortex pairs. Specifically, the leading vortex pairs up with the secondary vortex from the previous stroke, and the secondary vortex pairs up with the leading vortex of the following stroke. One may observe the high vortex structure similarity between the $f = 0.64$ Hz cases with flexible and rigid panels (FIG. \ref{vorticity rigid}(a) \& \ref{vorticity flexible}(a)). This is understandable since the amount of chord-wise bending reduces with lower $f$. Thus, the kinematics and shapes of both flexible and rigid panels at this $f$ are similar (FIG. \ref{A_TE}(c)), resulting in overall similar flow patterns (FIG. \ref{mean velocity rigid}(a) \& \ref{mean velocity flexible}(a)). At $f = 0.89$ Hz (FIG. \ref{vorticity flexible}(b)), a 2P+2S wake occurs due to the further severing of the TE vortex sheet. During every stroke, the long TE vortex splits into three relatively equal strength and spacing vortex cores. The two vortex cores on the side pair up with those from the previous and the next strokes. The vortex core in the middle sheds along the centerline alone as a single vortex. The vortex pair orients the jet in a direction that opposes the background mean flow, effectively slowing down $u$ along the outer regions of the wake (FIG. \ref{mean velocity flexible}(b)). The single vortices along the centerline interact with the vortex pairs on the side to generate more oscillatory mean $u$ profiles in the central wake region (FIG. \ref{mean velocity flexible}(b)). At $f = 1.00$ Hz (FIG. \ref{vorticity flexible}(c)), a bifurcating 2P wake diverges from the centerline of the pitching motion. The orientation of the vortex pair changes such that the jet has a slightly opposing angle initially and rotates to a neutral or slightly favorable angle with respect to the oncoming $u$ later downstream. Intuitively, the time-averaged velocity shows mild $u$ decrease close to the maximum positions of TE and mild increment along two bifurcating jets later downstream. Between the bifurcating jets is a small uniform convection zone with reduced $u$ (FIG. \ref{mean velocity flexible}(c)). At $f = 1.14$ Hz (FIG. \ref{vorticity flexible}(d)), the similar bifurcating 2P wake occurs. The orientation of the vortex pair directs the jet at a favorable angle which accelerates the oncoming $u$. Between the laterally parting vortex pairs, water moves slowly with weaker periodicity. Combined, the time-averaged velocity field has laterally bifurcating jets of increased $u$ and a central region with reduced $u$ for steady uniform thermal convection (FIG. \ref{mean velocity flexible}(d)). Despite the minor differences, $f = 1.00$ and $1.14$ Hz share the same vortex breakup mechanism. During every stroke, a TE vortex keeps accumulating circulation and stretches across the entire $A_{\mathrm{TE}}$. Close to the end of the stroke, this long vortex sheet severs into two, one weaker and one stronger. The weaker vortex pairs up with the stronger vortex from the previous stroke and gets shed. The stronger vortex lingers and pairs up with the weaker vortex from the next stroke. This 2P mechanism is fundamentally different from the 2P wake behind an oscillating non-perforated rigid foil reported by Schnipper \textit{et al.} \cite{schnipper2009vortex}. In their case, the 2P wake is created due to the time difference between the TE vortex creation and the LE vortex rolling across the entire chord. Thus, their vortex pair is consisted of one TE vortex and one LE vortex. However, in our case, the perforation prevents the roll and growth of the LE vortex due to the absence of the traditionally defined velocity boundary layer. Thus, both vortices in a pair come from the TE motion. 

    \subsection{Lagrangian coherent structures}

    Lagrangian coherent structures (LCS) have been widely used to understand the fluid mixing in ocean engineering \cite{lekien2005pollution}, turbulence \cite{mathur2007uncovering}, and biological flows \cite{peng2009transport, wu2024coherent}. Specifically, the hyperbolic LCS outlines the boundaries of the fluid pockets that undergo translation, rotation, and deformation over time. The Finite-Time Lyapunov Exponent (FTLE) is used to quantitatively illustrate the stable and unstable manifolds that physically correspond to fluid element contraction and stretching along the LCS respectively \cite{haller2015lagrangian}. The contracting and stretching of fluid elements in different directions creates saddle points that function as the expressway intersections for material transport and mixing. The mathematical theory for computing FTLE has been detailed in the pioneering works by Shadden, Haller, and Brunton, et al. \cite{shadden2005definition, haller2000lagrangian, brunton2010fast}
    
    We borrow the MATLAB package developed by Dabiri \textit{et al.} for computing the hyperbolic LCS. Foundations, validations, and applications of this package can be found in published works \cite{shadden2006lagrangian, peng2009transport}. Essentially, a matrix of equally-spacing particle flow map (our case: $194\times168$ particles) is initialized in the fluid velocity field obtained from the experiment (our case: phase-averaged velocity) and allowed to transport freely with the flow by an integration over forward- or backward-time. Across each time step from $t$ to $t+\delta t$, the flow map Jacobian $\mathbf{J}$ for each particle can be approximated as
    
    \begin{equation}\label{eq:Jacobian}
    \mathbf{J} = \begin{bmatrix} 
    \frac{\partial x(t_0+\delta t)}{\partial x(t_0)} & \frac{\partial x(t_0+\delta t)}{\partial y(t_0)} \\ 
    \frac{\partial y(t_0+\delta t)}{\partial x(t_0)} & \frac{\partial y(t_0+\delta t)}{\partial y(t_0)}
    \end{bmatrix} \, .
    \end{equation}

    
    \noindent The Cauchy-Green strain tensor ($\mathbf{\Delta}$) is equal to the multiplication of $\mathbf{J}$ with its own transpose $\mathbf{J}^\mathrm{T}$.The eigenvector and the eigenvalue of $\mathbf{\Delta}$ represent the direction and the amount of maximum stretching of the specific fluid element in forward- or backward-time depending on the integration. Then the FTLE can be calculated as 

    \begin{equation}\label{eq:FTLE}
    \mathrm{FTLE} = \frac{1}{|\delta t|}\mathrm{log}\left(\sqrt{\lambda_\mathrm{max}(\mathbf{\Delta})}\right)\, .
    \end{equation}
    
    \noindent where $\lambda_\mathrm{max}$ is the maximum eigenvalue of the matrix $\mathbf{\Delta}$. The FTLE from forward- or backward-time integration can be denoted as pFTLE or nFTLE respectively. 
    
    \begin{figure}[htbp]
    \centering
    \includegraphics[width=1\textwidth]{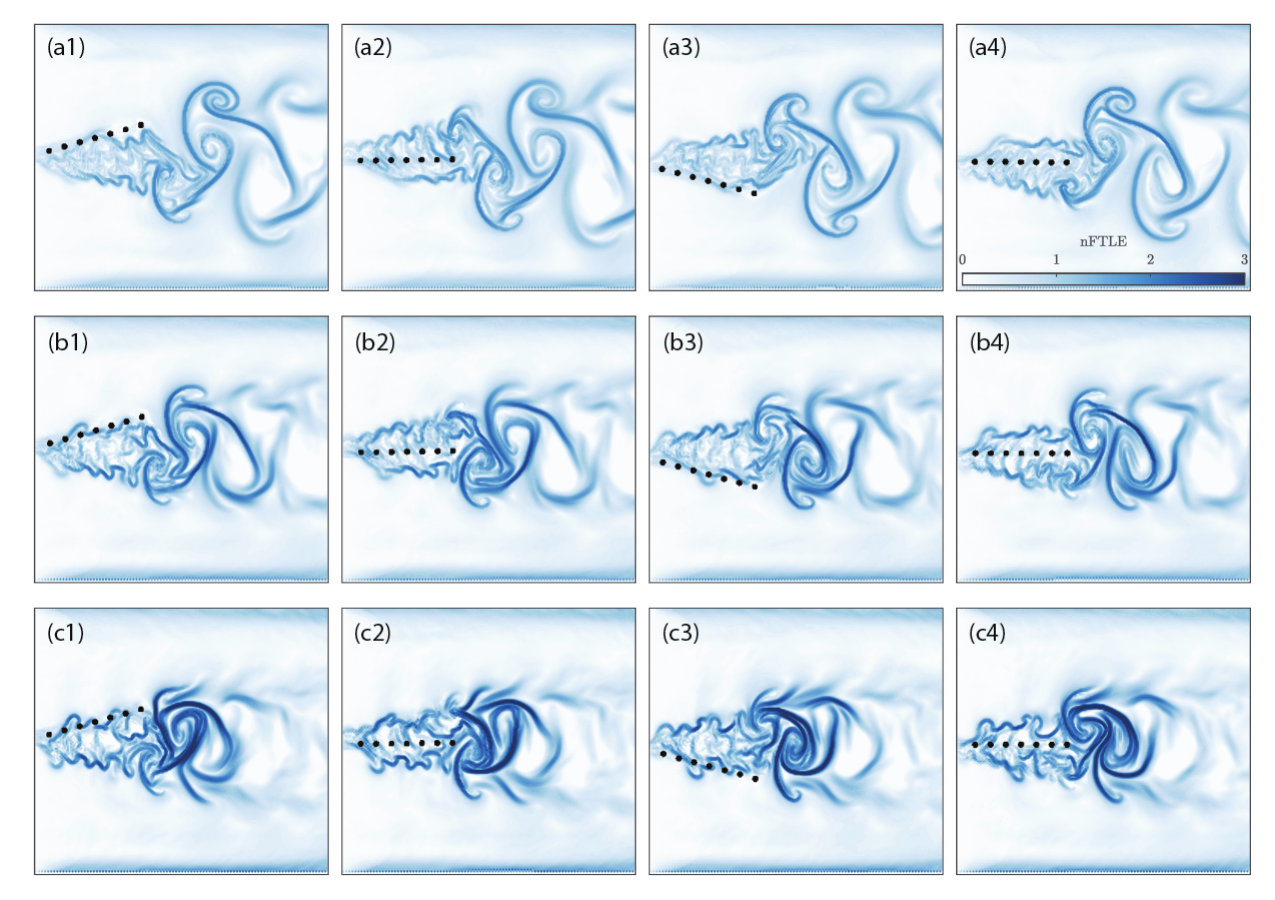}
    \caption{\textbf{Attractive LCS illustrated by nFTLE around the rigid panel.} (a) $f = 0.64$ Hz, (b) $f = 0.89$ Hz, (c) $f = 1.14$ Hz, (1-4) Numbers correspond to 0, 25, 50, 75\% progress of a phase-averaged cycle.} 
    \label{nFTLE rigid}
    \end{figure}
    
    In this work, we only compute and examine the nFTLE by integrating backward in $t$ for one stroke cycle. High nFTLE values form ridges that represent the unstable manifolds which express the flow structure stability since fluid elements are attracted to them, hence the other name ``attractive LCS''. Previous works have shown its ability to trace the movement of fluid tracers (chemical \cite{voth2002experimental} and inertial particles \cite{wu2024coherent}) in the flow, analogous to the thermal convection of hot water blobs. The hot water zone near the pitching panel can be thought as the fluid tracer. It will be attracted to the nFTLE ridges and stretched along these ridges to be mixed in the vortex wake. The shape of the hot water blobs will conform to the shape of the nFTLE ridges without much area change due to the incompressible approximation and 2D assumption. Therefore, the nFTLE ridges represent where the hot water will go. The shape of the nFTLE ridges shows us how the hot water blob will be stretched and whether it will fold onto itself \cite{haller2000lagrangian}. The distribution of the nFTLE field will tell us whether the hot water is stretched evenly across the domain or staying local. A wider distribution of higher nFTLE values means the hot water blob will be spread more quickly and widely, which encourages thermal convection and mixing. 
    
\begin{figure}[htbp]
    \centering
    \includegraphics[width=1\textwidth]{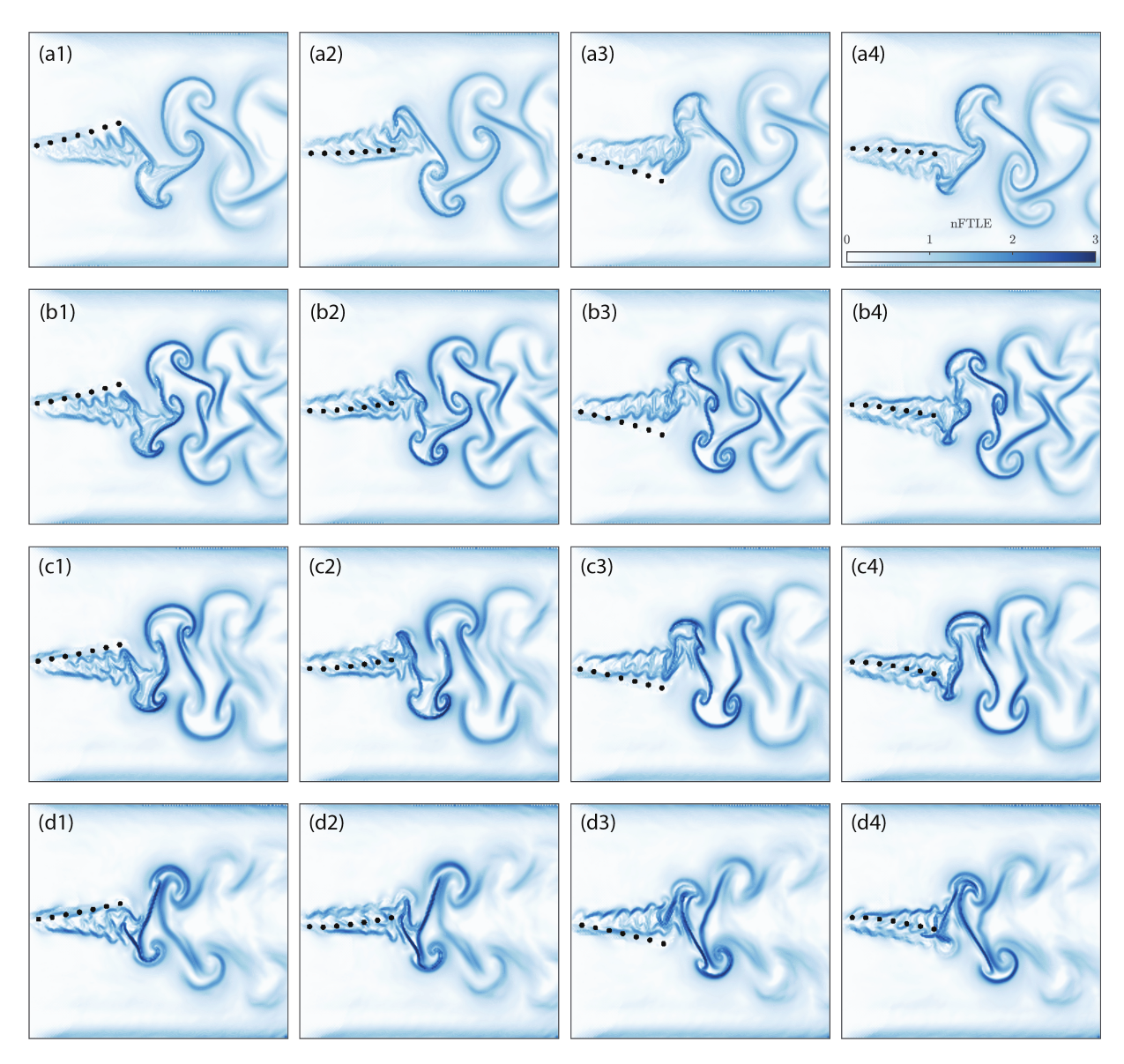}
    \caption{\textbf{Attractive LCS illustrated by nFTLE around the flexible panel.} (a) $f = 0.64$ Hz, (b) $f = 0.89$ Hz, (c) $f = 1.00$ Hz, (d) $f = 1.14$ Hz, (1-4) Numbers correspond to 0, 25, 50, 75\% progress of a phase-averaged cycle.} 
    \label{nFTLE flexible}
    \end{figure}

    As shown in FIG. \ref{nFTLE rigid} \& \ref{nFTLE flexible}, the ridges highlight the LCS in the wake to be consistent with the boundaries of the vortices shown in FIG. \ref{vorticity rigid} \& \ref{vorticity flexible}. On the superficial level, the nFTLE ridges confirm the identified vortex structure transitions from non-bifurcating 2P to 2P+2S then to bifurcating 2P with the flexible panel. On the fluid mixing perspective, the ridges function as an invisible boundary for fluid motion, meaning that fluid elements should remain on either side of the ridge without crossing over \cite{haller2015lagrangian, voth2002experimental}. This helps determine which fluid region comes from the hot zone swept by the perforated panel and which is entrained from the cold background flow on the side. 

    For the rigid panel, at $f = 0.64$ Hz (FIG. \ref{nFTLE rigid}(a)), the 2P wake sheds coherently along the centerline downstream, and there's enough separation between adjacent LCS to entrain cold water into the hotter wake to enhance mixing. As $f$ increases to 0.89 and 1.14 Hz (FIG. \ref{nFTLE rigid}(b-c)), the nFTLE ridges are highly compacted right behind the TE with relatively high values. This represents a strong stretching and folding of the hot water limited to this localized area without much persistence into the farther regions. The continuous bounding shear layers mentioned before for $f = 1.14$ Hz also match the nFTLE ridges on the side that provide nearly no entrance for cold water entrainment. From this trend, we anticipate that pitching the rigid panel at a higher frequency would negatively influence thermal mixing in the wake. 
    
    For the flexible panel, at $f = 0.64$ Hz and $f = 0.89$ Hz (FIG. \ref{nFTLE flexible}(a-b)), large flow entrainment entrances exist which allow the lateral thermal mixing between the hot and cool water to happen more aggressively in the wake region. Note the LCSs for both panels are very similar at $f = 0.64$ Hz, as discussed already in Section IV.C. As $f$ increases to 1.00 and 1.14 Hz (FIG. 
    \ref{nFTLE flexible}(c-d)), the laterally growing LCS (mushroom-shaped pockets corresponding to the vortex pairs) from different stroke cycles are relatively close to each other, making the nFTLE ridges compacted. This leaves the entrance for lateral flow entrainment to be relatively smaller, although still much larger than the rigid panel at the same frequencies. Therefore, we anticipate the thermal mixing performance to be overall better with the flexible panel than the rigid panel. 

    \begin{figure}[htbp]
    \centering
    \includegraphics[width=1\textwidth]{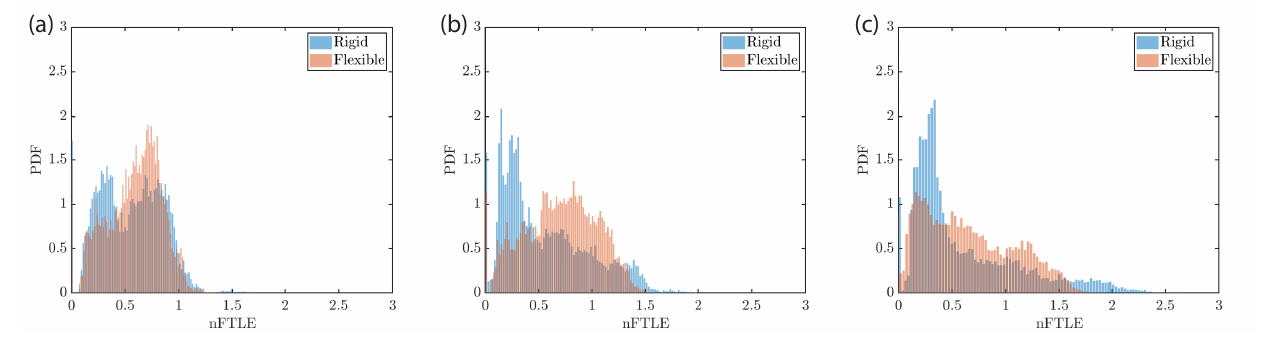}
    \caption{\textbf{Probability density function (PDF) of the time-averaged nFTLE over one stroke cycle in the wake region.} Comparisons were done between rigid and flexible panels for three representative pitching frequencies, (a) $f = 0.64$ Hz, (b) $f = 0.89$ Hz, and (c) $f = 1.14$ Hz. } 
    \label{nFTLE PDF}
    \end{figure}
    
\begin{figure}[htbp]
    \centering
    \includegraphics[width=1\textwidth]{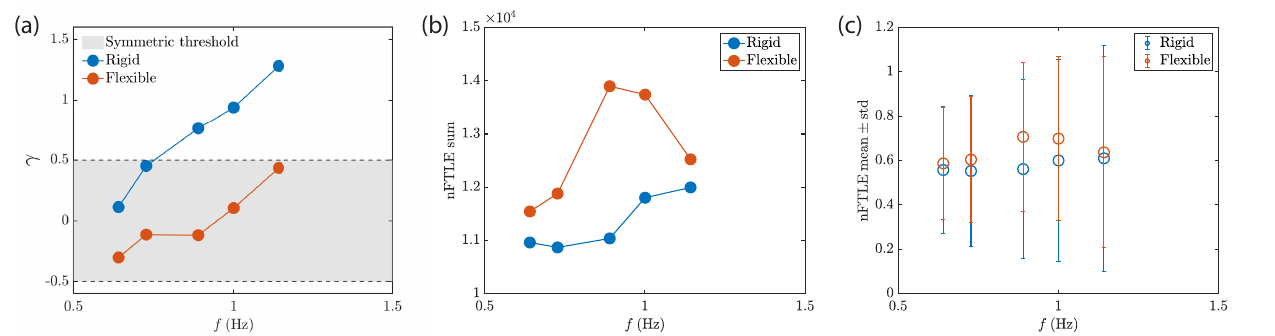}
    \caption{\textbf{Statistical analysis of the nFTLE field.} (a) Skewness of the nFTLE distributions in the region of interest ($x>0.035$ m) at $f = $ 0.64, 0.73, 0.89, 1.00, and 1.14 Hz. (b) Sum of the nFTLE values representing total fluid stretching in the same cases. (c) Mean and standard deviation of the nFTLE values representing the spread of fluid stretching in the same cases. }
    \label{nFTLE stats}
    \end{figure}

    Statistical analysis can be done on the nFTLE in the wake region beyond the panel ($x>0.035$ m) to quantify the stretching distribution and anticipate different thermal mixing performances. Probability density functions (PDF) of the mean nFTLE field over one stroke cycle help us visualize the distribution of fluid stretching in the region of interest. As shown in FIG. \ref{nFTLE PDF}, the flexible panel generally has a higher nFTLE contribution in the ranges of $0.5\sim 0.8$ (FIG. \ref{nFTLE PDF}(a)), $0.5\sim 1.2$ (FIG. \ref{nFTLE PDF}(b)), and $0.5\sim 1.5$ (FIG. \ref{nFTLE PDF}(c)). This indicates that the flexible panel's motion creates stronger fluid stretching of the hot water blob in a wider region in the wake, which is helpful for dispensing the heat in a larger range. One can notice that at $f = 0.89$ and $1.14$ Hz, the rigid panel has a longer right-skewed tail with high nFTLE values that are absent for the flexible panel. This is in consistency with the nFTLE contours showing that the rigid panel can induce stronger fluid stretching than the flexible panel, but only in a very localized fashion. The right-skewed high nFTLE values indeed lead to some chaotic mixing \cite{liu1994quantification} and enhance the thermal mixing near the TE, but does not translate to the overall mixing effect. To further quantify the nFTLE distribution, we compute the skewness ($\gamma$), sum, mean, and standard deviation within the wake region. Regarding skewness, homogeneous and heterogeneous mixing have been defined to separate the cases of global uniform mixing and strong localized mixing \cite{badas2017quantification, beron2010surface}. For homogeneous mixing, the nFTLE distribution is expected to follow a Gaussian distribution ($\gamma = 0$), while heterogeneous mixing's would be highly skewed. Empirically, a skewness value $\gamma\in[-0.5,0.5]$ can be classified as approximately symmetric, in other words more homogeneous mixing. The flexible panel fell within the symmetric threshold for all tested $f$, whereas the rigid panel fell in the symmetric threshold only at $f \in [0.64,0.73]$ Hz (FIG. \ref{nFTLE stats}(a)). Specifically, the rigid panel continues to become more skewed with an increasing $f$, whereas the flexible panel has less skewness for an intermediate $f$ range. However, we refrain from making further claims due to $\gamma$'s sensitivity to experimental values. We argue that the distinction between inside and outside the symmetric threshold has more practical implications on mixing. We also found that the flexible panel yields a generally higher sum and mean of nFTLE in the wake region, indicating its better thermal mixing ability (FIG. \ref{nFTLE stats}(b-c)). 

    \subsection{Thermal mixing of the temperature field}

\begin{figure}[htbp]
    \centering
    \includegraphics[width=1\textwidth]{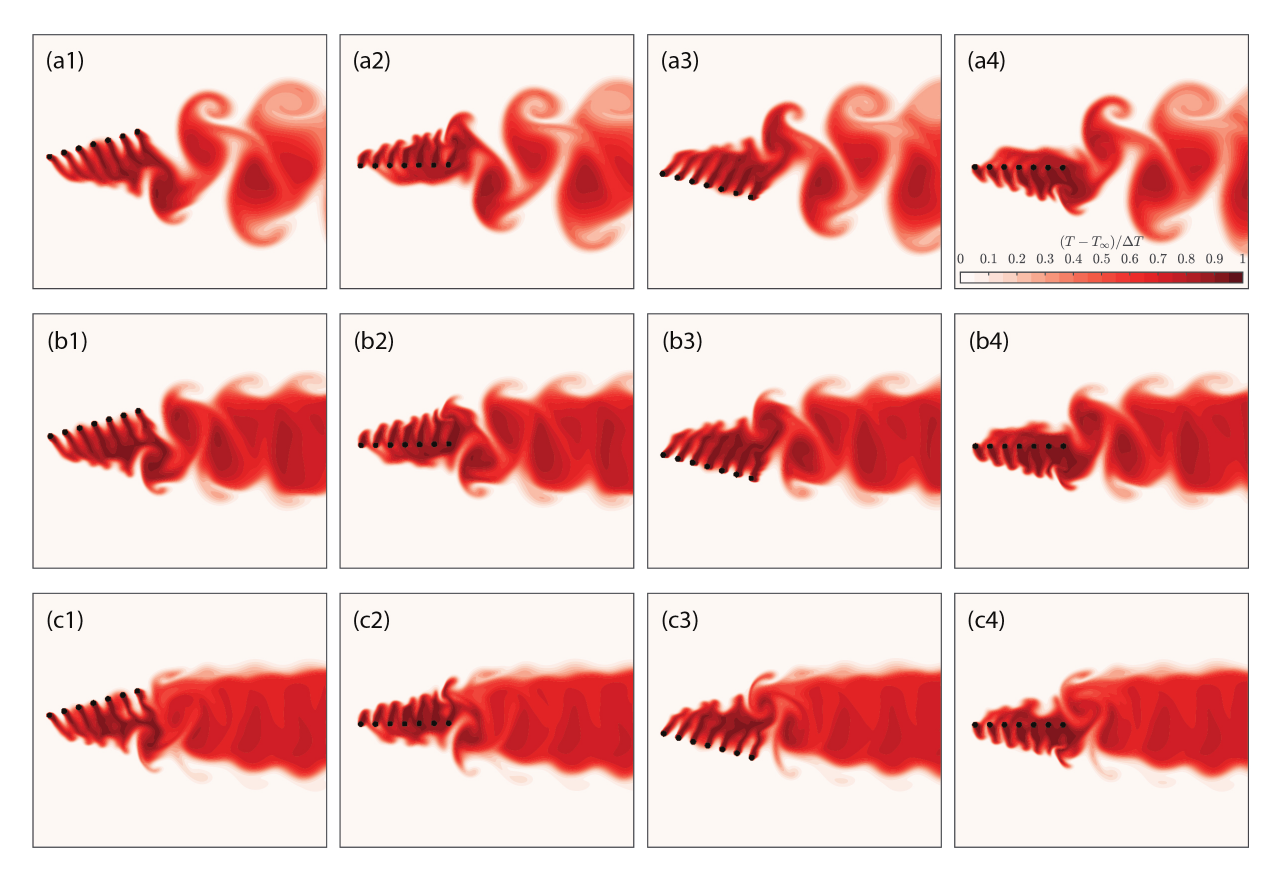}
    \caption{\textbf{Temperature fields during the 11th stroke cycle of the rigid panel.} (a) $f = 0.64$ Hz, (b) $f = 0.89$ Hz, (c) $f = 1.14$ Hz, (1-4) Numbers correspond to 0, 25, 50, 75\% progress of a phase-averaged cycle.} 
    \label{T rigid}
    \end{figure}

\begin{figure}[htbp]
    \centering
    \includegraphics[width=1\textwidth]{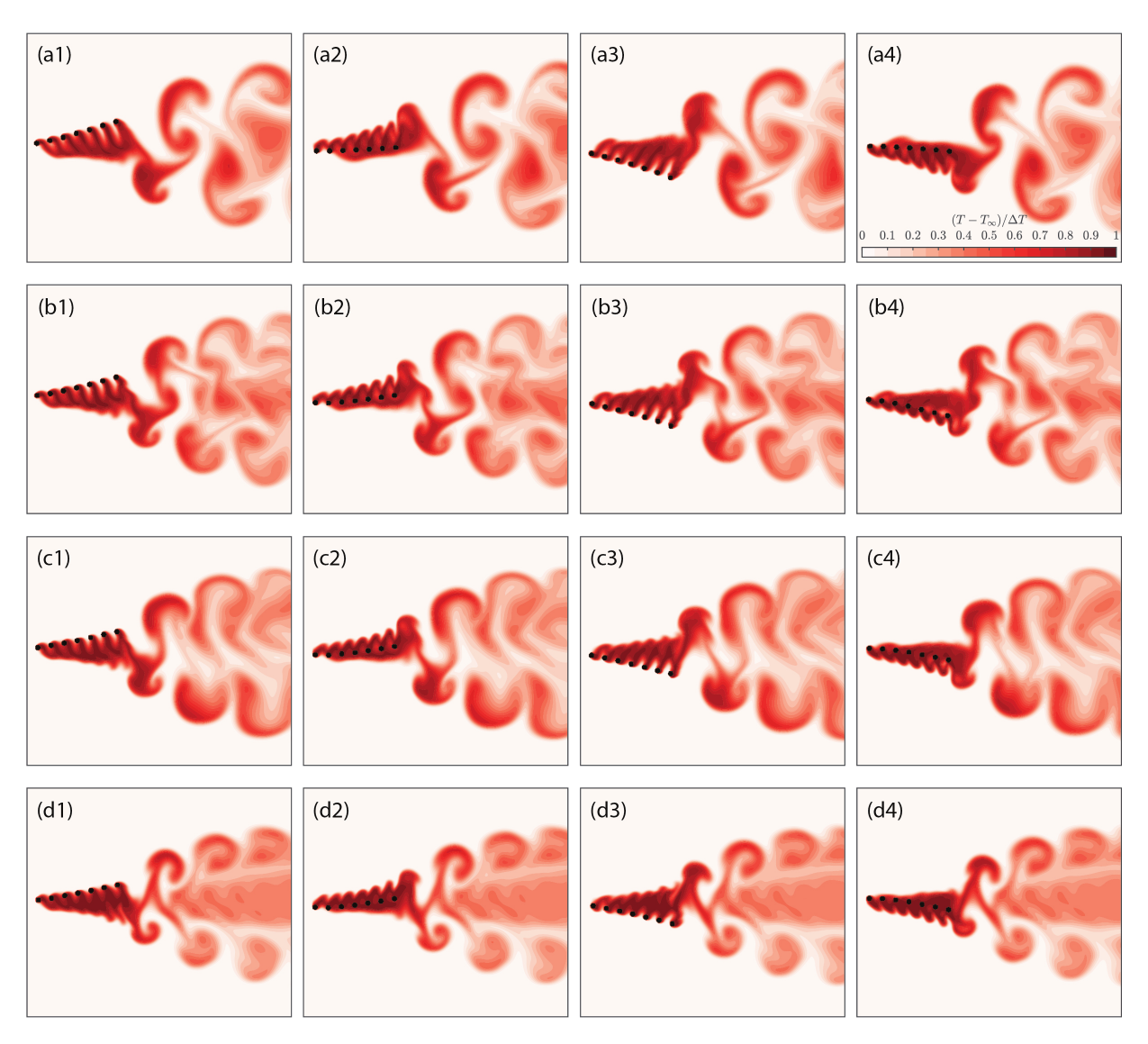}
    \caption{\textbf{Temperature fields during the 11th stroke cycle of the flexible panel.} (a) $f = 0.64$ Hz, (b) $f = 0.89$ Hz, (c) $f = 1.00$ Hz, (d) $f = 1.14$ Hz, (1-4) Numbers correspond to 0, 25, 50, 75\% progress of a phase-averaged cycle.} 
    \label{T flexible}
    \end{figure}

    Finally, we examine the simulated temperature field to consolidate the understandings from the previous analysis. Since the uniform background inflow is assumed to have a constantly low temperature source and all heating comes from the solid posts separating the perforation holes on the panel, the mixing performance should be characterized by the spread of $T$ in the domain. As expected, the high temperature fluid pockets leave the thermal boundary layer next to the panel and gets transported into the wake with stretching and cold fluid entrainment paths outlined by the nFTLE ridges around the vortex boundaries (FIG. \ref{T rigid} \& \ref{T flexible}). This is consistent with the definition of attractive LCS as fluid boundaries and unstable manifolds. 

\begin{figure}[htbp]
    \centering
    \includegraphics[width=1\textwidth]{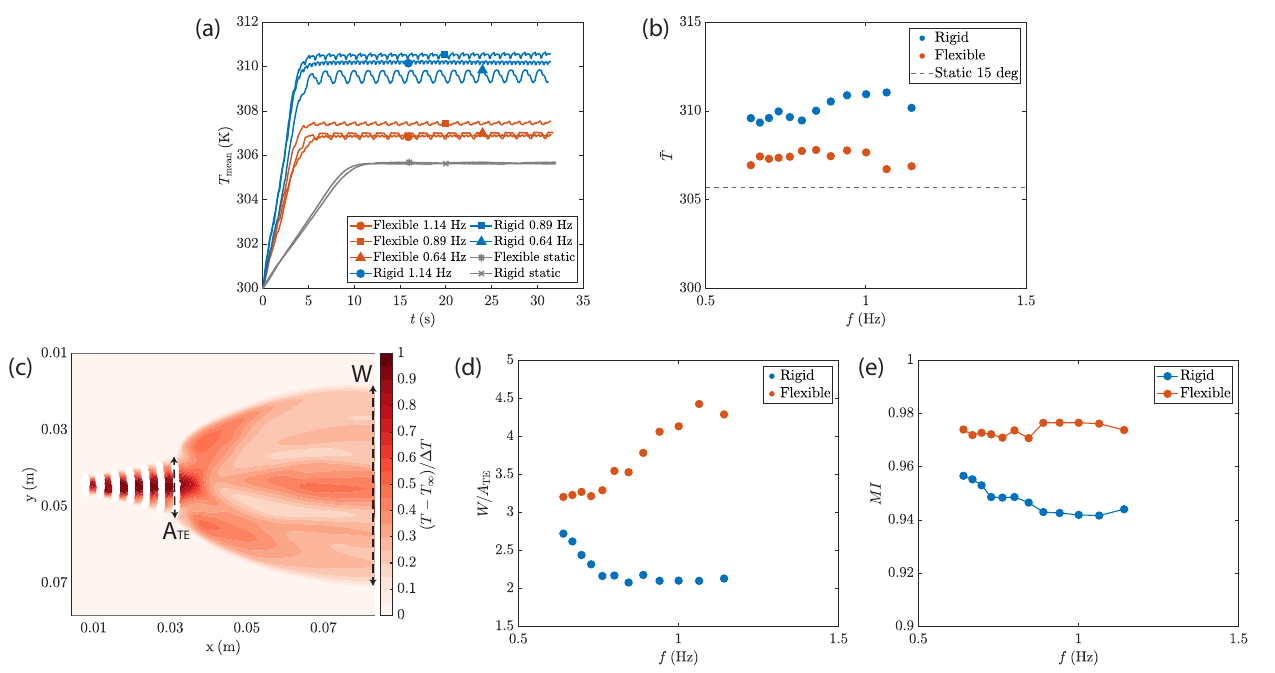}
    \caption{\textbf{Equilibrium temperature field over the last 5 pitching cycles of the simulation.} (a) Domain-averaged temperature evolution over time from example cases. (b) Equilibrium temperatures for all tested conditions. (c) Definition of the thermal wake width. (d) Normalized thermal wake width as a function of $f$. (e) Mixing index at the right outlet.} 
    \label{T field stats}
    \end{figure}

    Comparing the $T_\mathrm{mean}$ evolution with time (FIG. \ref{T field stats}(a)) and the $\bar{T}$ across the last 5 pitching cycles (FIG. \ref{T field stats}(b)) from the simulation, both rigid and flexible panels performed better than the static case [Section IV.I], proving that the prescribed motion is effectively enhancing the thermal performance. Counterintuitively, the rigid panel reaches a higher $\bar{T}$ than the flexible panel in all cases. This is likely from the effect of a short-lived and localized LCS in the wake, which does not contribute to streamwise convection of heat as much as the flexible panel wake (compare FIG. \ref{mean velocity rigid} \& \ref{mean velocity flexible}). The shear layer blockage of the lateral entrainment also adds to the excessive heating in the rigid panel wake. If the purpose is to elevate the $\bar{T}$ of the domain, the rigid panel may be a better option than using flexible material. However, $\bar{T}$ alone does not represent the thermal mixing capability, which could be more important for certain applications where a more uniform downstream temperature distribution is needed. The thermal wake width to TE amplitude ratio ($W/A_\mathrm{TE}$) reflects the panel's ability to transfer heat in the transverse direction. Results in FIG. \ref{T field stats}(c-d) shows that the flexible panel is indeed more capable of distributing heat in a wider thermal wake. The true $W$ of the flexible panel stays relatively constant with increasing $f$, but the deflection keeps decreasing the $A_\mathrm{TE}$, which leads to an increasing $W/A_\mathrm{TE}$. This means for the same flexible panel, pitching at a higher $f$ is able to maintain the same level of lateral thermal dispersion with a smaller lateral movement. Since $A_\mathrm{TE}$ remains a constant for the rigid panel, the $W/A_\mathrm{TE}$ trend reflects the true $W$. As $f$ increases, a shrinkage of the thermal wake width is noticed for 0.64 - 0.76 Hz, then followed by a plateau at higher $f$. This indicates that the rigid panel's capability to transfer heat laterally in the wake does not benefit from faster pitching, opposite from the flexible panel. Another metric to evaluate thermal mixing is the equilibrium outlet mixing index ($MI$), defined as

    \begin{equation}\label{eq:MI}
    MI = \frac{1}{\Delta t}\int_{t_0}^{t_0+\Delta t} \left(1-\frac{1}{T_{\mathrm{mean}}^{\mathrm{outlet}}(t)}\sqrt{\frac{\sum_{i = 1}^{n} \left[T(y_i,t)-T_{\mathrm{mean}}^\mathrm{outlet}(t)\right]^2}{n}} \right) dt\, ,
    \end{equation}
    
    \noindent following the work of Lambert and Rangel \cite{lambert2010role}, where $\Delta t$ (5 cycles) is the time-averaging window size, $T(y_i, t)$ is the local temperature value at various $y$ locations along the outlet column, $T_\mathrm{mean}^\mathrm{outlet}$ is the mean temperature across the outlet column, and $n$ is the number of data points in the outlet column. $MI$ is by definition a value between 0 (no mixing) and 1 (fully mixed). As shown in FIG. \ref{T field stats}(e), the flexible panel achieved overall better mixing at the outlet than the rigid panel. As mentioned before, the flexible panel's thermal wake width $W$ does not change significantly as a function of $f$, and thus the outlet $MI$ stays relatively constant across all $f$. However, the rigid panel did show a $W$ shrinkage with increasing $f$ in the lower range, which is reflected as the decreasing $MI$ for the same range. Note that faster pitching is compensated by potentially more power input due to the faster solid object motion and higher hydrodynamic power. Quantification of the power input requires a deeper understanding of the flexural rigidity, hydrodynamic forcing, and added mass effect of such perforated panels. Since the current focus is on the vortex transition and thermal mixing, we leave efficiency evaluation for future exploration. 

\begin{figure}[htbp]
    \centering
    \includegraphics[width=1\textwidth]{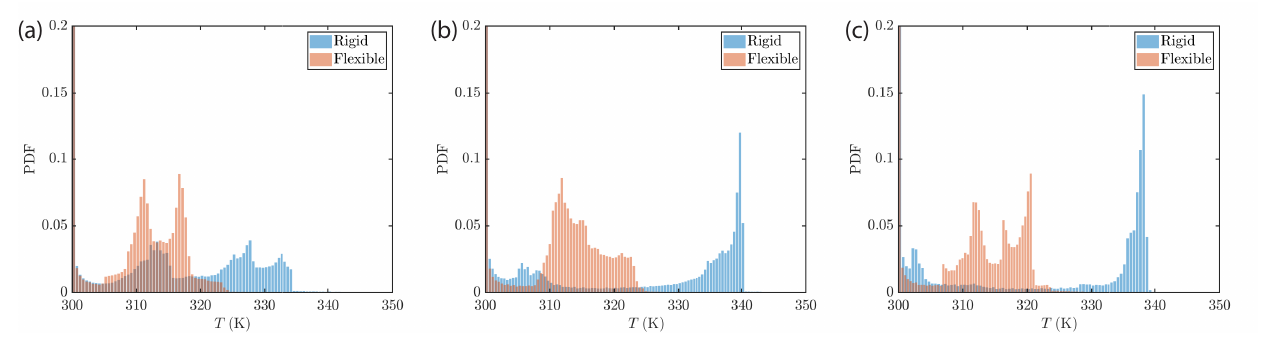}
    \caption{\textbf{Temperature contributions in the time-averaged field over the last 5 pitching cycles in the simulation for the wake region beyond the panel ($x>0.035\:\mathrm{m}$).} Comparisons quantify the different thermal mixing and temperature distributions between the rigid and flexible panels at (a) $f = 0.64$ Hz, (b) $f = 0.89$ Hz, and (c) $f = 1.14$ Hz. } 
    \label{T PDF}
    \end{figure}

    Similar to the nFTLE field, we also examine the probability density function of the wake $T$ field ($x>0.035$ m) averaged over the last five cycles to distinguish the performances between rigid and flexible panels. Overall, the rigid panel yielded more higher temperature concentrations in the field than the flexible panel, also shown in FIG. \ref{T field stats}(a-b). At low $f$, the rigid panel benefits from the vortex structure transition and the periodic breaking of the bounding shear layer, hence the resulting $T$ distribution in FIG. \ref{T PDF}(a). At higher $f$, the rigid panel results in one spike at 300 K, one spike at around 335 - 340 K, and nearly nothing in between (FIG. \ref{T PDF}(b-c)). This indicates low thermal mixing due to the sharp delineation between the interior and exterior sides of the wake. The non-mixing wake separates the resulting temperature field into an almost uniformly heated wake and non-heated external flow (heterogeneous thermal mixing). On the other hand, the flexible panel across all $f$ resulted in better thermal mixing. A wider distribution in the intermediate $T$ range demonstrate the more uniform distribution of heat throughout the flow field (homogeneous thermal mixing). 

\section{Conclusion and Discussion}
    Our work provides the first characterization of actively pitching perforated panels regarding the vortex structure transition, Lagrangian coherent structures, and thermal mixing. Comparisons were made between panels with different chord-wise flexibilities and pitching frequencies. Our results demonstrated that the thermal mixing benefits greatly from the prescribed motion due to the creation of unsteady wake vortex streets. Panels with pitching motion, regardless of the chord-wise flexibility and pitching frequency, all reached a higher domain-averaged equilibrium temperature than the static scenario by an approximately 30\% - 78\% margin (FIG. \ref{T field stats}(a-b)). The panel without chord-wise flexibility does not experience a fundamental vortex structure change within the tested $f$ range. Lateral entrainment and the associated thermal mixing between the hotter wake flow and the colder incoming flow depend on the periodic breaking of the bounding shear layers which becomes prominent only at lower frequency ranges. Due to local vortex annihilation, thermal convection is suppressed in the wake, resulting in a sharper temperature gradient across the bounding shear layers. The panel with chord-wise flexibility, however, experiences fundamentally different vortex structure modes when pitched at different $f$. Specifically, a transition from non-bifurcating 2P to 2P+2S then to bifurcating 2P is documented along with increasing frequency. The perforation invokes a different vortex shedding mechanism than those described on non-perforated hydrofoils \cite{schnipper2009vortex}. Both LCS and simulated temperature showed that chord-wise flexibility enhances lateral thermal mixing entrainment and more-homogeneous mixing. Our work innovatively explores the FSI of actuated flexible panels with porosity, a topic that is very much neglected in the literature and underestimated for its application. We drew new connections between unsteady hydrodynamics of vortex shedding and heat-mass transfer. Our work not only extends the fundamental understanding of flexible vortex generators but also provides insights for novel biomimetic designs of thermal and material dispensing systems. Several directions for future work are discussed below.
    
    \subsection{Simulation boundary conditions}
    In order to study the vortex generation and thermal mixing, the current Convection Diffusion simulation adopts a simplified boundary condition on both the fluid and the solid domain. The moving solid structure is isothermal at 350 K. When the solid structure moves to the new location in the next frame, the ghost cells in the fluid domain (the nodes previously occupied by solid) are assumed to be heated up to 350 K as the initial condition for the next time segment's simulation [Details explained in Section III.B.]. This essentially assumes that we have an ultra-effective heat source that allows infinitely fast thermal conduction of heat from the hotter solid to the surrounding fluid. In reality, the amount of conductive heating would depend on the temperature gradient, thermal conductivity, and time, making it possible for the ghost cell fluid elements to start at a temperature lower than the solid's. The same issue applies to the material transport problems as the cross-membrane filtration depends on the concentration gradient, diffusion coefficient, membrane porosity, and time. Currently, we make sense of our assumptions due to the fine temporal discretizations from the high-speed video and the ODE45 auto-stepping: large overlapping of the moving solid structure exists in two adjacent frames. The Neumann constant flux boundary condition at the flow outlet may also introduce error. Though common in similar channel flow simulations \cite{saini2023performance, jin2024enhancing, park2017simulation}, the horizontal domain size may be too small to assume complete stabilization of the flow and temperature in the streamwise direction, owing to the periodic wake. For now, we are comfortable with our simulation results since the solver validations match published data. However, we acknowledge the limitations of this current simulation method. More realistic settings of the boundary conditions (such like isothermal, isoflux, and more complicated cases depending on the physical properties) and iterative time stepping schemes might improve the simulation to reflect more application-relevant engineering situations. Such improvement will enable us to study local transport phenomenon near the perforations including saturation and fatigue \cite{lou2024wing}.

    \subsection{Complex actuation}
    Other kinematics and flexural rigidities are worth investigation too. As discussed by Quinn \textit{et al.} \cite{quinn2014scaling}, flexible panels' deformation in fluid can be modelled by the Euler-Bernoulli Bean Theory, which predicts different modes of resonance frequencies. As a function of the kinematic boundary condition (sinusoidal pitching, heaving, flapping, temporal asymmetry, and spatial asymmetry) and the flexural rigidity, these resonance modes may play a role in the vortex formation and thermal mixing, as they do to hydrodynamic force production \cite{quinn2014scaling, liu2022fine}. We conducted preliminary experiments with LE pitching at even higher frequencies ($\sim 3$ Hz). The flexible panel sustains more deformation, similar to the second resonance. The panel, though slender and perforated, acts like a bluff body and creates a non-shedding circulating wake like those observed by Shang \textit{et al.} \cite{shang2018flow}. This wake has an oscillation frequency much slower than the pitching $f$, making it trickier for the simulation pre-processing. Due to the lack of strong vortex-induced mixing, we believe that they do not have better thermal mixing efficacy. Yet, explaining the FSI could contribute to the overall understanding of actuated perforated panels. 
    
    \subsection{Implications to ichthyology and biomimetic engineering}
    Echoing with the biological inspiration from fish gill filaments, we believe future extensions from the present work may shed lights on the evolutionary optimization of this respiratory organ. Fish gills are subject to different incoming flow conditions: continuous for ram ventilation and pulsatile for opercular pumping \cite{hughes1972morphometrics}. Fish gills also operate at a wide range of $Re$, making the optimal pore size important for gas exchange \cite{park2014optimal}. Furthermore, fish have multiple gill sets in parallel, potentially harvesting benefits from the inter-lamellar interactions. All these details can be translated to similar engineering problems, so the exploration potential is vast.

\section{Acknowledgement}
We are grateful for the discussions with Haibo Dong, Zihao Huang, and Daniel B. Quinn at the University of Virginia, and the Bio-inspired Fluid Lab at Cornell University. We thank Casey Dillman at Cornell University Museum of Vertebrates for providing fish gill samples. Y.F. acknowledges the funding support from Cornell University CALS Alumni Associate Award. Y.F. and S.J. acknowledge the partial funding support from the U.S. National Oceanic and Atmospheric Administration (NOAA-NA24OARX417C0598-T1-01) and the U.S. National Science Foundation (CMMI-2042740 \& CBET-2401507). 

\section{Author contributions}

Y.F. conceived the idea; Y.F., S.T., and J.G. conducted experiments. Y.F. and Z.L. conducted simulations. Y.F., S.T., and J.G. analyzed data. S.J. supervised the research. Y.F., S.T., and J.G. wrote the original manuscript. Y.F., Z.L., S.T., J.G., and S.J. reviewed the manuscript.

\section{Conflict of interests}

The authors declare no competing interests.

\newpage


\begin{thebibliography}{92}%
\makeatletter
\providecommand \@ifxundefined [1]{%
 \@ifx{#1\undefined}
}%
\providecommand \@ifnum [1]{%
 \ifnum #1\expandafter \@firstoftwo
 \else \expandafter \@secondoftwo
 \fi
}%
\providecommand \@ifx [1]{%
 \ifx #1\expandafter \@firstoftwo
 \else \expandafter \@secondoftwo
 \fi
}%
\providecommand \natexlab [1]{#1}%
\providecommand \enquote  [1]{``#1''}%
\providecommand \bibnamefont  [1]{#1}%
\providecommand \bibfnamefont [1]{#1}%
\providecommand \citenamefont [1]{#1}%
\providecommand \href@noop [0]{\@secondoftwo}%
\providecommand \href [0]{\begingroup \@sanitize@url \@href}%
\providecommand \@href[1]{\@@startlink{#1}\@@href}%
\providecommand \@@href[1]{\endgroup#1\@@endlink}%
\providecommand \@sanitize@url [0]{\catcode `\\12\catcode `\$12\catcode `\&12\catcode `\#12\catcode `\^12\catcode `\_12\catcode `\%12\relax}%
\providecommand \@@startlink[1]{}%
\providecommand \@@endlink[0]{}%
\providecommand \url  [0]{\begingroup\@sanitize@url \@url }%
\providecommand \@url [1]{\endgroup\@href {#1}{\urlprefix }}%
\providecommand \urlprefix  [0]{URL }%
\providecommand \Eprint [0]{\href }%
\providecommand \doibase [0]{http://dx.doi.org/}%
\providecommand \selectlanguage [0]{\@gobble}%
\providecommand \bibinfo  [0]{\@secondoftwo}%
\providecommand \bibfield  [0]{\@secondoftwo}%
\providecommand \translation [1]{[#1]}%
\providecommand \BibitemOpen [0]{}%
\providecommand \bibitemStop [0]{}%
\providecommand \bibitemNoStop [0]{.\EOS\space}%
\providecommand \EOS [0]{\spacefactor3000\relax}%
\providecommand \BibitemShut  [1]{\csname bibitem#1\endcsname}%
\let\auto@bib@innerbib\@empty
\bibitem [{\citenamefont {Tong}\ and\ \citenamefont {Elimelech}(2016)}]{tong2016global}%
  \BibitemOpen
  \bibfield  {author} {\bibinfo {author} {\bibfnamefont {Tiezheng}\ \bibnamefont {Tong}}\ and\ \bibinfo {author} {\bibfnamefont {Menachem}\ \bibnamefont {Elimelech}},\ }\bibfield  {title} {\enquote {\bibinfo {title} {The global rise of zero liquid discharge for wastewater management: drivers, technologies, and future directions},}\ }\href@noop {} {\bibfield  {journal} {\bibinfo  {journal} {Environmental science \& technology}\ }\textbf {\bibinfo {volume} {50}},\ \bibinfo {pages} {6846--6855} (\bibinfo {year} {2016})}\BibitemShut {NoStop}%
\bibitem [{\citenamefont {Vikas}\ and\ \citenamefont {Dwarakish}(2015)}]{vikas2015coastal}%
  \BibitemOpen
  \bibfield  {author} {\bibinfo {author} {\bibfnamefont {Mangalore}\ \bibnamefont {Vikas}}\ and\ \bibinfo {author} {\bibfnamefont {GS}~\bibnamefont {Dwarakish}},\ }\bibfield  {title} {\enquote {\bibinfo {title} {Coastal pollution: a review},}\ }\href@noop {} {\bibfield  {journal} {\bibinfo  {journal} {Aquatic procedia}\ }\textbf {\bibinfo {volume} {4}},\ \bibinfo {pages} {381--388} (\bibinfo {year} {2015})}\BibitemShut {NoStop}%
\bibitem [{\citenamefont {Khalaj}\ and\ \citenamefont {Halgamuge}(2017)}]{khalaj2017review}%
  \BibitemOpen
  \bibfield  {author} {\bibinfo {author} {\bibfnamefont {Ali~Habibi}\ \bibnamefont {Khalaj}}\ and\ \bibinfo {author} {\bibfnamefont {Saman~K}\ \bibnamefont {Halgamuge}},\ }\bibfield  {title} {\enquote {\bibinfo {title} {A review on efficient thermal management of air-and liquid-cooled data centers: From chip to the cooling system},}\ }\href@noop {} {\bibfield  {journal} {\bibinfo  {journal} {Applied energy}\ }\textbf {\bibinfo {volume} {205}},\ \bibinfo {pages} {1165--1188} (\bibinfo {year} {2017})}\BibitemShut {NoStop}%
\bibitem [{\citenamefont {Wu}\ and\ \citenamefont {Wang}(2006)}]{wu2006combined}%
  \BibitemOpen
  \bibfield  {author} {\bibinfo {author} {\bibfnamefont {DaWei}\ \bibnamefont {Wu}}\ and\ \bibinfo {author} {\bibfnamefont {RuZhu}\ \bibnamefont {Wang}},\ }\bibfield  {title} {\enquote {\bibinfo {title} {Combined cooling, heating and power: A review},}\ }\href@noop {} {\bibfield  {journal} {\bibinfo  {journal} {progress in energy and combustion science}\ }\textbf {\bibinfo {volume} {32}},\ \bibinfo {pages} {459--495} (\bibinfo {year} {2006})}\BibitemShut {NoStop}%
\bibitem [{\citenamefont {Himmelfarb}\ \emph {et~al.}(2020)\citenamefont {Himmelfarb}, \citenamefont {Vanholder}, \citenamefont {Mehrotra},\ and\ \citenamefont {Tonelli}}]{himmelfarb2020current}%
  \BibitemOpen
  \bibfield  {author} {\bibinfo {author} {\bibfnamefont {Jonathan}\ \bibnamefont {Himmelfarb}}, \bibinfo {author} {\bibfnamefont {Raymond}\ \bibnamefont {Vanholder}}, \bibinfo {author} {\bibfnamefont {Rajnish}\ \bibnamefont {Mehrotra}}, \ and\ \bibinfo {author} {\bibfnamefont {Marcello}\ \bibnamefont {Tonelli}},\ }\bibfield  {title} {\enquote {\bibinfo {title} {The current and future landscape of dialysis},}\ }\href@noop {} {\bibfield  {journal} {\bibinfo  {journal} {Nature Reviews Nephrology}\ }\textbf {\bibinfo {volume} {16}},\ \bibinfo {pages} {573--585} (\bibinfo {year} {2020})}\BibitemShut {NoStop}%
\bibitem [{\citenamefont {Wood}\ and\ \citenamefont {Eom}(2021)}]{wood2021osmorespiratory}%
  \BibitemOpen
  \bibfield  {author} {\bibinfo {author} {\bibfnamefont {Chris~M}\ \bibnamefont {Wood}}\ and\ \bibinfo {author} {\bibfnamefont {Junho}\ \bibnamefont {Eom}},\ }\bibfield  {title} {\enquote {\bibinfo {title} {The osmorespiratory compromise in the fish gill},}\ }\href@noop {} {\bibfield  {journal} {\bibinfo  {journal} {Comparative Biochemistry and Physiology Part A: Molecular \& Integrative Physiology}\ }\textbf {\bibinfo {volume} {254}},\ \bibinfo {pages} {110895} (\bibinfo {year} {2021})}\BibitemShut {NoStop}%
\bibitem [{\citenamefont {Hughes}\ \emph {et~al.}(2010)\citenamefont {Hughes}, \citenamefont {Price},\ and\ \citenamefont {Banks}}]{hughes2010predators}%
  \BibitemOpen
  \bibfield  {author} {\bibinfo {author} {\bibfnamefont {Nelika~K}\ \bibnamefont {Hughes}}, \bibinfo {author} {\bibfnamefont {Catherine~J}\ \bibnamefont {Price}}, \ and\ \bibinfo {author} {\bibfnamefont {Peter~B}\ \bibnamefont {Banks}},\ }\bibfield  {title} {\enquote {\bibinfo {title} {Predators are attracted to the olfactory signals of prey},}\ }\href@noop {} {\bibfield  {journal} {\bibinfo  {journal} {PLoS One}\ }\textbf {\bibinfo {volume} {5}},\ \bibinfo {pages} {e13114} (\bibinfo {year} {2010})}\BibitemShut {NoStop}%
\bibitem [{\citenamefont {Wisenden}(2000)}]{wisenden2000olfactory}%
  \BibitemOpen
  \bibfield  {author} {\bibinfo {author} {\bibfnamefont {Brain~D}\ \bibnamefont {Wisenden}},\ }\bibfield  {title} {\enquote {\bibinfo {title} {Olfactory assessment of predation risk in the aquatic environment},}\ }\href@noop {} {\bibfield  {journal} {\bibinfo  {journal} {Philosophical Transactions of the Royal Society of London. Series B: Biological Sciences}\ }\textbf {\bibinfo {volume} {355}},\ \bibinfo {pages} {1205--1208} (\bibinfo {year} {2000})}\BibitemShut {NoStop}%
\bibitem [{\citenamefont {Reddy}\ \emph {et~al.}(2022)\citenamefont {Reddy}, \citenamefont {Murthy},\ and\ \citenamefont {Vergassola}}]{reddy2022olfactory}%
  \BibitemOpen
  \bibfield  {author} {\bibinfo {author} {\bibfnamefont {Gautam}\ \bibnamefont {Reddy}}, \bibinfo {author} {\bibfnamefont {Venkatesh~N}\ \bibnamefont {Murthy}}, \ and\ \bibinfo {author} {\bibfnamefont {Massimo}\ \bibnamefont {Vergassola}},\ }\bibfield  {title} {\enquote {\bibinfo {title} {Olfactory sensing and navigation in turbulent environments},}\ }\href@noop {} {\bibfield  {journal} {\bibinfo  {journal} {Annual Review of Condensed Matter Physics}\ }\textbf {\bibinfo {volume} {13}},\ \bibinfo {pages} {191--213} (\bibinfo {year} {2022})}\BibitemShut {NoStop}%
\bibitem [{\citenamefont {Chen}\ and\ \citenamefont {Haviland-Jones}(2000)}]{chen2000human}%
  \BibitemOpen
  \bibfield  {author} {\bibinfo {author} {\bibfnamefont {Denise}\ \bibnamefont {Chen}}\ and\ \bibinfo {author} {\bibfnamefont {Jeannette}\ \bibnamefont {Haviland-Jones}},\ }\bibfield  {title} {\enquote {\bibinfo {title} {Human olfactory communication of emotion},}\ }\href@noop {} {\bibfield  {journal} {\bibinfo  {journal} {Perceptual and motor skills}\ }\textbf {\bibinfo {volume} {91}},\ \bibinfo {pages} {771--781} (\bibinfo {year} {2000})}\BibitemShut {NoStop}%
\bibitem [{\citenamefont {Grieves}\ \emph {et~al.}(2022)\citenamefont {Grieves}, \citenamefont {Gilles}, \citenamefont {Cuthill}, \citenamefont {Sz{\'e}kely}, \citenamefont {MacDougall-Shackleton},\ and\ \citenamefont {Caspers}}]{grieves2022olfactory}%
  \BibitemOpen
  \bibfield  {author} {\bibinfo {author} {\bibfnamefont {Leanne~A}\ \bibnamefont {Grieves}}, \bibinfo {author} {\bibfnamefont {Marc}\ \bibnamefont {Gilles}}, \bibinfo {author} {\bibfnamefont {Innes~C}\ \bibnamefont {Cuthill}}, \bibinfo {author} {\bibfnamefont {Tam{\'a}s}\ \bibnamefont {Sz{\'e}kely}}, \bibinfo {author} {\bibfnamefont {Elizabeth~A}\ \bibnamefont {MacDougall-Shackleton}}, \ and\ \bibinfo {author} {\bibfnamefont {Barbara~A}\ \bibnamefont {Caspers}},\ }\bibfield  {title} {\enquote {\bibinfo {title} {Olfactory camouflage and communication in birds},}\ }\href@noop {} {\bibfield  {journal} {\bibinfo  {journal} {Biological Reviews}\ }\textbf {\bibinfo {volume} {97}},\ \bibinfo {pages} {1193--1209} (\bibinfo {year} {2022})}\BibitemShut {NoStop}%
\bibitem [{\citenamefont {Aref}(2002)}]{aref2002development}%
  \BibitemOpen
  \bibfield  {author} {\bibinfo {author} {\bibfnamefont {Hassan}\ \bibnamefont {Aref}},\ }\bibfield  {title} {\enquote {\bibinfo {title} {The development of chaotic advection},}\ }\href@noop {} {\bibfield  {journal} {\bibinfo  {journal} {Physics of fluids}\ }\textbf {\bibinfo {volume} {14}},\ \bibinfo {pages} {1315--1325} (\bibinfo {year} {2002})}\BibitemShut {NoStop}%
\bibitem [{\citenamefont {Aref}(2020)}]{aref2020stirring}%
  \BibitemOpen
  \bibfield  {author} {\bibinfo {author} {\bibfnamefont {Hassan}\ \bibnamefont {Aref}},\ }\bibfield  {title} {\enquote {\bibinfo {title} {Stirring by chaotic advection},}\ }in\ \href@noop {} {\emph {\bibinfo {booktitle} {Hamiltonian Dynamical Systems}}}\ (\bibinfo  {publisher} {CRC Press},\ \bibinfo {year} {2020})\ pp.\ \bibinfo {pages} {725--745}\BibitemShut {NoStop}%
\bibitem [{\citenamefont {Dimotakis}(2005)}]{dimotakis2005turbulent}%
  \BibitemOpen
  \bibfield  {author} {\bibinfo {author} {\bibfnamefont {Paul~E}\ \bibnamefont {Dimotakis}},\ }\bibfield  {title} {\enquote {\bibinfo {title} {Turbulent mixing},}\ }\href@noop {} {\bibfield  {journal} {\bibinfo  {journal} {Annu. Rev. Fluid Mech.}\ }\textbf {\bibinfo {volume} {37}},\ \bibinfo {pages} {329--356} (\bibinfo {year} {2005})}\BibitemShut {NoStop}%
\bibitem [{\citenamefont {Caulfield}(2021)}]{caulfield2021layering}%
  \BibitemOpen
  \bibfield  {author} {\bibinfo {author} {\bibfnamefont {CP}~\bibnamefont {Caulfield}},\ }\bibfield  {title} {\enquote {\bibinfo {title} {Layering, instabilities, and mixing in turbulent stratified flows},}\ }\href@noop {} {\bibfield  {journal} {\bibinfo  {journal} {Annual Review of Fluid Mechanics}\ }\textbf {\bibinfo {volume} {53}},\ \bibinfo {pages} {113--145} (\bibinfo {year} {2021})}\BibitemShut {NoStop}%
\bibitem [{\citenamefont {Zhang}\ \emph {et~al.}(2021)\citenamefont {Zhang}, \citenamefont {Liu}, \citenamefont {Sund{\'e}n},\ and\ \citenamefont {Xie}}]{zhang2021combined}%
  \BibitemOpen
  \bibfield  {author} {\bibinfo {author} {\bibfnamefont {Guohua}\ \bibnamefont {Zhang}}, \bibinfo {author} {\bibfnamefont {Jian}\ \bibnamefont {Liu}}, \bibinfo {author} {\bibfnamefont {Bengt}\ \bibnamefont {Sund{\'e}n}}, \ and\ \bibinfo {author} {\bibfnamefont {Gongnan}\ \bibnamefont {Xie}},\ }\bibfield  {title} {\enquote {\bibinfo {title} {Combined experimental and numerical studies on flow characteristic and heat transfer in ribbed channels with vortex generators of various types and arrangements},}\ }\href@noop {} {\bibfield  {journal} {\bibinfo  {journal} {International Journal of Thermal Sciences}\ }\textbf {\bibinfo {volume} {167}},\ \bibinfo {pages} {107036} (\bibinfo {year} {2021})}\BibitemShut {NoStop}%
\bibitem [{\citenamefont {Luo}\ \emph {et~al.}(2016)\citenamefont {Luo}, \citenamefont {Wen}, \citenamefont {Wang}, \citenamefont {Sund{\'e}n},\ and\ \citenamefont {Wang}}]{luo2016thermal}%
  \BibitemOpen
  \bibfield  {author} {\bibinfo {author} {\bibfnamefont {Lei}\ \bibnamefont {Luo}}, \bibinfo {author} {\bibfnamefont {Fengbo}\ \bibnamefont {Wen}}, \bibinfo {author} {\bibfnamefont {Lei}\ \bibnamefont {Wang}}, \bibinfo {author} {\bibfnamefont {Bengt}\ \bibnamefont {Sund{\'e}n}}, \ and\ \bibinfo {author} {\bibfnamefont {Songtao}\ \bibnamefont {Wang}},\ }\bibfield  {title} {\enquote {\bibinfo {title} {Thermal enhancement by using grooves and ribs combined with delta-winglet vortex generator in a solar receiver heat exchanger},}\ }\href@noop {} {\bibfield  {journal} {\bibinfo  {journal} {Applied energy}\ }\textbf {\bibinfo {volume} {183}},\ \bibinfo {pages} {1317--1332} (\bibinfo {year} {2016})}\BibitemShut {NoStop}%
\bibitem [{\citenamefont {Promvonge}\ \emph {et~al.}(2010)\citenamefont {Promvonge}, \citenamefont {Chompookham}, \citenamefont {Kwankaomeng},\ and\ \citenamefont {Thianpong}}]{promvonge2010enhanced}%
  \BibitemOpen
  \bibfield  {author} {\bibinfo {author} {\bibfnamefont {Pongjet}\ \bibnamefont {Promvonge}}, \bibinfo {author} {\bibfnamefont {Teerapat}\ \bibnamefont {Chompookham}}, \bibinfo {author} {\bibfnamefont {Sutapat}\ \bibnamefont {Kwankaomeng}}, \ and\ \bibinfo {author} {\bibfnamefont {Chinaruk}\ \bibnamefont {Thianpong}},\ }\bibfield  {title} {\enquote {\bibinfo {title} {Enhanced heat transfer in a triangular ribbed channel with longitudinal vortex generators},}\ }\href@noop {} {\bibfield  {journal} {\bibinfo  {journal} {Energy Conversion and Management}\ }\textbf {\bibinfo {volume} {51}},\ \bibinfo {pages} {1242--1249} (\bibinfo {year} {2010})}\BibitemShut {NoStop}%
\bibitem [{\citenamefont {He}\ \emph {et~al.}(2021)\citenamefont {He}, \citenamefont {Deng},\ and\ \citenamefont {Feng}}]{he2021film}%
  \BibitemOpen
  \bibfield  {author} {\bibinfo {author} {\bibfnamefont {Juan}\ \bibnamefont {He}}, \bibinfo {author} {\bibfnamefont {Qinghua}\ \bibnamefont {Deng}}, \ and\ \bibinfo {author} {\bibfnamefont {Zhenping}\ \bibnamefont {Feng}},\ }\bibfield  {title} {\enquote {\bibinfo {title} {Film cooling performance enhancement by upstream v-shaped protrusion/dimple vortex generator},}\ }\href@noop {} {\bibfield  {journal} {\bibinfo  {journal} {International Journal of Heat and Mass Transfer}\ }\textbf {\bibinfo {volume} {180}},\ \bibinfo {pages} {121784} (\bibinfo {year} {2021})}\BibitemShut {NoStop}%
\bibitem [{\citenamefont {Wang}\ \emph {et~al.}(2015)\citenamefont {Wang}, \citenamefont {Chen}, \citenamefont {Liaw},\ and\ \citenamefont {Tseng}}]{wang2015experimental}%
  \BibitemOpen
  \bibfield  {author} {\bibinfo {author} {\bibfnamefont {Chi-Chuan}\ \bibnamefont {Wang}}, \bibinfo {author} {\bibfnamefont {Kuan-Yu}\ \bibnamefont {Chen}}, \bibinfo {author} {\bibfnamefont {Jane-Sunn}\ \bibnamefont {Liaw}}, \ and\ \bibinfo {author} {\bibfnamefont {Chih-Yung}\ \bibnamefont {Tseng}},\ }\bibfield  {title} {\enquote {\bibinfo {title} {An experimental study of the air-side performance of fin-and-tube heat exchangers having plain, louver, and semi-dimple vortex generator configuration},}\ }\href@noop {} {\bibfield  {journal} {\bibinfo  {journal} {International Journal of Heat and Mass Transfer}\ }\textbf {\bibinfo {volume} {80}},\ \bibinfo {pages} {281--287} (\bibinfo {year} {2015})}\BibitemShut {NoStop}%
\bibitem [{\citenamefont {Tang}\ \emph {et~al.}(2016)\citenamefont {Tang}, \citenamefont {Chu}, \citenamefont {Ahmed},\ and\ \citenamefont {Zeng}}]{tang2016new}%
  \BibitemOpen
  \bibfield  {author} {\bibinfo {author} {\bibfnamefont {LH}~\bibnamefont {Tang}}, \bibinfo {author} {\bibfnamefont {WX}~\bibnamefont {Chu}}, \bibinfo {author} {\bibfnamefont {N}~\bibnamefont {Ahmed}}, \ and\ \bibinfo {author} {\bibfnamefont {M}~\bibnamefont {Zeng}},\ }\bibfield  {title} {\enquote {\bibinfo {title} {A new configuration of winglet longitudinal vortex generator to enhance heat transfer in a rectangular channel},}\ }\href@noop {} {\bibfield  {journal} {\bibinfo  {journal} {Applied Thermal Engineering}\ }\textbf {\bibinfo {volume} {104}},\ \bibinfo {pages} {74--84} (\bibinfo {year} {2016})}\BibitemShut {NoStop}%
\bibitem [{\citenamefont {Han}\ \emph {et~al.}(2024)\citenamefont {Han}, \citenamefont {Li},\ and\ \citenamefont {Zhang}}]{han2024insight}%
  \BibitemOpen
  \bibfield  {author} {\bibinfo {author} {\bibfnamefont {Wenbo}\ \bibnamefont {Han}}, \bibinfo {author} {\bibfnamefont {Wei}\ \bibnamefont {Li}}, \ and\ \bibinfo {author} {\bibfnamefont {Hongpeng}\ \bibnamefont {Zhang}},\ }\bibfield  {title} {\enquote {\bibinfo {title} {Insight into mixing performance of bionic fractal baffle micromixers based on murray's law},}\ }\href@noop {} {\bibfield  {journal} {\bibinfo  {journal} {International Communications in Heat and Mass Transfer}\ }\textbf {\bibinfo {volume} {157}},\ \bibinfo {pages} {107843} (\bibinfo {year} {2024})}\BibitemShut {NoStop}%
\bibitem [{\citenamefont {Liao}\ and\ \citenamefont {Jing}(2023)}]{liao2023experimental}%
  \BibitemOpen
  \bibfield  {author} {\bibinfo {author} {\bibfnamefont {Wei}\ \bibnamefont {Liao}}\ and\ \bibinfo {author} {\bibfnamefont {Dalei}\ \bibnamefont {Jing}},\ }\bibfield  {title} {\enquote {\bibinfo {title} {Experimental study on fluid mixing and pressure drop of mini-mixer with flexible vortex generator},}\ }\href@noop {} {\bibfield  {journal} {\bibinfo  {journal} {International Communications in Heat and Mass Transfer}\ }\textbf {\bibinfo {volume} {142}},\ \bibinfo {pages} {106615} (\bibinfo {year} {2023})}\BibitemShut {NoStop}%
\bibitem [{\citenamefont {Zhong}\ \emph {et~al.}(2022)\citenamefont {Zhong}, \citenamefont {Fu}, \citenamefont {Chan}, \citenamefont {Yang}, \citenamefont {Qiu},\ and\ \citenamefont {Chao}}]{zhong2022experimental}%
  \BibitemOpen
  \bibfield  {author} {\bibinfo {author} {\bibfnamefont {XL}~\bibnamefont {Zhong}}, \bibinfo {author} {\bibfnamefont {Sau~Chung}\ \bibnamefont {Fu}}, \bibinfo {author} {\bibfnamefont {Ka~Chung}\ \bibnamefont {Chan}}, \bibinfo {author} {\bibfnamefont {G}~\bibnamefont {Yang}}, \bibinfo {author} {\bibfnamefont {HH}~\bibnamefont {Qiu}}, \ and\ \bibinfo {author} {\bibfnamefont {Christopher~YH}\ \bibnamefont {Chao}},\ }\bibfield  {title} {\enquote {\bibinfo {title} {Experimental study on the thermal-hydraulic performance of a fluttering split flag in a channel flow},}\ }\href@noop {} {\bibfield  {journal} {\bibinfo  {journal} {International Journal of Heat and Mass Transfer}\ }\textbf {\bibinfo {volume} {182}},\ \bibinfo {pages} {121945} (\bibinfo {year} {2022})}\BibitemShut {NoStop}%
\bibitem [{\citenamefont {Liu}\ \emph {et~al.}(2024)\citenamefont {Liu}, \citenamefont {Leng},\ and\ \citenamefont {Wang}}]{liu2024heat}%
  \BibitemOpen
  \bibfield  {author} {\bibinfo {author} {\bibfnamefont {Xueling}\ \bibnamefont {Liu}}, \bibinfo {author} {\bibfnamefont {Yunkai}\ \bibnamefont {Leng}}, \ and\ \bibinfo {author} {\bibfnamefont {Jiansheng}\ \bibnamefont {Wang}},\ }\bibfield  {title} {\enquote {\bibinfo {title} {The heat transfer enhancement with a flag-shaped flexible wing},}\ }\href@noop {} {\bibfield  {journal} {\bibinfo  {journal} {International Journal of Heat and Mass Transfer}\ }\textbf {\bibinfo {volume} {224}},\ \bibinfo {pages} {125362} (\bibinfo {year} {2024})}\BibitemShut {NoStop}%
\bibitem [{\citenamefont {Lee}\ \emph {et~al.}(2017)\citenamefont {Lee}, \citenamefont {Park}, \citenamefont {Kim}, \citenamefont {Ryu},\ and\ \citenamefont {Sung}}]{lee2017heat}%
  \BibitemOpen
  \bibfield  {author} {\bibinfo {author} {\bibfnamefont {Jae~Bok}\ \bibnamefont {Lee}}, \bibinfo {author} {\bibfnamefont {Sung~Goon}\ \bibnamefont {Park}}, \bibinfo {author} {\bibfnamefont {Boyoung}\ \bibnamefont {Kim}}, \bibinfo {author} {\bibfnamefont {Jaeha}\ \bibnamefont {Ryu}}, \ and\ \bibinfo {author} {\bibfnamefont {Hyung~Jin}\ \bibnamefont {Sung}},\ }\bibfield  {title} {\enquote {\bibinfo {title} {Heat transfer enhancement by flexible flags clamped vertically in a poiseuille channel flow},}\ }\href@noop {} {\bibfield  {journal} {\bibinfo  {journal} {International Journal of Heat and Mass Transfer}\ }\textbf {\bibinfo {volume} {107}},\ \bibinfo {pages} {391--402} (\bibinfo {year} {2017})}\BibitemShut {NoStop}%
\bibitem [{\citenamefont {Jin}\ \emph {et~al.}(2024)\citenamefont {Jin}, \citenamefont {Leng}, \citenamefont {Wang}, \citenamefont {Zhang},\ and\ \citenamefont {Cui}}]{jin2024enhancing}%
  \BibitemOpen
  \bibfield  {author} {\bibinfo {author} {\bibfnamefont {Yuzhen}\ \bibnamefont {Jin}}, \bibinfo {author} {\bibfnamefont {Chunhui}\ \bibnamefont {Leng}}, \bibinfo {author} {\bibfnamefont {Zhaokun}\ \bibnamefont {Wang}}, \bibinfo {author} {\bibfnamefont {Xuming}\ \bibnamefont {Zhang}}, \ and\ \bibinfo {author} {\bibfnamefont {Jingyu}\ \bibnamefont {Cui}},\ }\bibfield  {title} {\enquote {\bibinfo {title} {Enhancing heat transfer in laminar channel flow by tuning the mass distribution of a flexible reed},}\ }\href@noop {} {\bibfield  {journal} {\bibinfo  {journal} {Physics of Fluids}\ }\textbf {\bibinfo {volume} {36}} (\bibinfo {year} {2024})}\BibitemShut {NoStop}%
\bibitem [{\citenamefont {Chen}\ \emph {et~al.}(2024)\citenamefont {Chen}, \citenamefont {Liu},\ and\ \citenamefont {Sung}}]{chen2024enhancement}%
  \BibitemOpen
  \bibfield  {author} {\bibinfo {author} {\bibfnamefont {Zepeng}\ \bibnamefont {Chen}}, \bibinfo {author} {\bibfnamefont {Yingzheng}\ \bibnamefont {Liu}}, \ and\ \bibinfo {author} {\bibfnamefont {Hyung~Jin}\ \bibnamefont {Sung}},\ }\bibfield  {title} {\enquote {\bibinfo {title} {Enhancement of heat transfer by a buckled flexible filament in a channel flow},}\ }\href@noop {} {\bibfield  {journal} {\bibinfo  {journal} {International Journal of Heat and Mass Transfer}\ }\textbf {\bibinfo {volume} {224}},\ \bibinfo {pages} {125364} (\bibinfo {year} {2024})}\BibitemShut {NoStop}%
\bibitem [{\citenamefont {Park}(2020)}]{park2020heat}%
  \BibitemOpen
  \bibfield  {author} {\bibinfo {author} {\bibfnamefont {Sung~Goon}\ \bibnamefont {Park}},\ }\bibfield  {title} {\enquote {\bibinfo {title} {Heat transfer enhancement by a wall-mounted flexible vortex generator with an inclination angle},}\ }\href@noop {} {\bibfield  {journal} {\bibinfo  {journal} {International Journal of Heat and Mass Transfer}\ }\textbf {\bibinfo {volume} {148}},\ \bibinfo {pages} {119053} (\bibinfo {year} {2020})}\BibitemShut {NoStop}%
\bibitem [{\citenamefont {Chen}\ \emph {et~al.}(2020)\citenamefont {Chen}, \citenamefont {Yang}, \citenamefont {Liu},\ and\ \citenamefont {Sung}}]{chen2020heat}%
  \BibitemOpen
  \bibfield  {author} {\bibinfo {author} {\bibfnamefont {Yujia}\ \bibnamefont {Chen}}, \bibinfo {author} {\bibfnamefont {Jongmin}\ \bibnamefont {Yang}}, \bibinfo {author} {\bibfnamefont {Yingzheng}\ \bibnamefont {Liu}}, \ and\ \bibinfo {author} {\bibfnamefont {Hyung~Jin}\ \bibnamefont {Sung}},\ }\bibfield  {title} {\enquote {\bibinfo {title} {Heat transfer enhancement in a poiseuille channel flow by using multiple wall-mounted flexible flags},}\ }\href@noop {} {\bibfield  {journal} {\bibinfo  {journal} {International Journal of Heat and Mass Transfer}\ }\textbf {\bibinfo {volume} {163}},\ \bibinfo {pages} {120447} (\bibinfo {year} {2020})}\BibitemShut {NoStop}%
\bibitem [{\citenamefont {Ali}\ \emph {et~al.}(2016)\citenamefont {Ali}, \citenamefont {Menanteau}, \citenamefont {Habchi}, \citenamefont {Lemenand},\ and\ \citenamefont {Harion}}]{ali2016heat}%
  \BibitemOpen
  \bibfield  {author} {\bibinfo {author} {\bibfnamefont {Samer}\ \bibnamefont {Ali}}, \bibinfo {author} {\bibfnamefont {S{\'e}bastien}\ \bibnamefont {Menanteau}}, \bibinfo {author} {\bibfnamefont {Charbel}\ \bibnamefont {Habchi}}, \bibinfo {author} {\bibfnamefont {Thierry}\ \bibnamefont {Lemenand}}, \ and\ \bibinfo {author} {\bibfnamefont {Jean-Luc}\ \bibnamefont {Harion}},\ }\bibfield  {title} {\enquote {\bibinfo {title} {Heat transfer and mixing enhancement by using multiple freely oscillating flexible vortex generators},}\ }\href@noop {} {\bibfield  {journal} {\bibinfo  {journal} {Applied Thermal Engineering}\ }\textbf {\bibinfo {volume} {105}},\ \bibinfo {pages} {276--289} (\bibinfo {year} {2016})}\BibitemShut {NoStop}%
\bibitem [{\citenamefont {Ding}\ \emph {et~al.}(2024)\citenamefont {Ding}, \citenamefont {Xiong}, \citenamefont {Han}, \citenamefont {Zhu},\ and\ \citenamefont {Ran}}]{ding2024improvement}%
  \BibitemOpen
  \bibfield  {author} {\bibinfo {author} {\bibfnamefont {Lin}\ \bibnamefont {Ding}}, \bibinfo {author} {\bibfnamefont {Jinzhen}\ \bibnamefont {Xiong}}, \bibinfo {author} {\bibfnamefont {Yuxiong}\ \bibnamefont {Han}}, \bibinfo {author} {\bibfnamefont {Zheyu}\ \bibnamefont {Zhu}}, \ and\ \bibinfo {author} {\bibfnamefont {Jingyu}\ \bibnamefont {Ran}},\ }\bibfield  {title} {\enquote {\bibinfo {title} {Improvement of mixing performance in laminar micromixers utilizing vortex-induced vibration of two circular cylinders},}\ }\href@noop {} {\bibfield  {journal} {\bibinfo  {journal} {International Communications in Heat and Mass Transfer}\ }\textbf {\bibinfo {volume} {159}},\ \bibinfo {pages} {108285} (\bibinfo {year} {2024})}\BibitemShut {NoStop}%
\bibitem [{\citenamefont {Zhang}\ \emph {et~al.}(2020)\citenamefont {Zhang}, \citenamefont {He},\ and\ \citenamefont {Zhang}}]{zhang2020fluid}%
  \BibitemOpen
  \bibfield  {author} {\bibinfo {author} {\bibfnamefont {Xiang}\ \bibnamefont {Zhang}}, \bibinfo {author} {\bibfnamefont {Guowei}\ \bibnamefont {He}}, \ and\ \bibinfo {author} {\bibfnamefont {Xing}\ \bibnamefont {Zhang}},\ }\bibfield  {title} {\enquote {\bibinfo {title} {Fluid--structure interactions of single and dual wall-mounted 2d flexible filaments in a laminar boundary layer},}\ }\href@noop {} {\bibfield  {journal} {\bibinfo  {journal} {Journal of Fluids and Structures}\ }\textbf {\bibinfo {volume} {92}},\ \bibinfo {pages} {102787} (\bibinfo {year} {2020})}\BibitemShut {NoStop}%
\bibitem [{\citenamefont {Hughes}(1972)}]{hughes1972morphometrics}%
  \BibitemOpen
  \bibfield  {author} {\bibinfo {author} {\bibfnamefont {George~M}\ \bibnamefont {Hughes}},\ }\bibfield  {title} {\enquote {\bibinfo {title} {Morphometrics of fish gills},}\ }\href@noop {} {\bibfield  {journal} {\bibinfo  {journal} {Respiration Physiology}\ }\textbf {\bibinfo {volume} {14}},\ \bibinfo {pages} {1--25} (\bibinfo {year} {1972})}\BibitemShut {NoStop}%
\bibitem [{\citenamefont {Li}\ \emph {et~al.}(2018)\citenamefont {Li}, \citenamefont {Dong},\ and\ \citenamefont {Zhao}}]{li2018balance}%
  \BibitemOpen
  \bibfield  {author} {\bibinfo {author} {\bibfnamefont {Chengyu}\ \bibnamefont {Li}}, \bibinfo {author} {\bibfnamefont {Haibo}\ \bibnamefont {Dong}}, \ and\ \bibinfo {author} {\bibfnamefont {Kai}\ \bibnamefont {Zhao}},\ }\bibfield  {title} {\enquote {\bibinfo {title} {A balance between aerodynamic and olfactory performance during flight in drosophila},}\ }\href@noop {} {\bibfield  {journal} {\bibinfo  {journal} {Nature communications}\ }\textbf {\bibinfo {volume} {9}},\ \bibinfo {pages} {3215} (\bibinfo {year} {2018})}\BibitemShut {NoStop}%
\bibitem [{\citenamefont {Lou}\ \emph {et~al.}(2024)\citenamefont {Lou}, \citenamefont {Lei}, \citenamefont {Dong},\ and\ \citenamefont {Li}}]{lou2024wing}%
  \BibitemOpen
  \bibfield  {author} {\bibinfo {author} {\bibfnamefont {Zhipeng}\ \bibnamefont {Lou}}, \bibinfo {author} {\bibfnamefont {Menglong}\ \bibnamefont {Lei}}, \bibinfo {author} {\bibfnamefont {Haibo}\ \bibnamefont {Dong}}, \ and\ \bibinfo {author} {\bibfnamefont {Chengyu}\ \bibnamefont {Li}},\ }\bibfield  {title} {\enquote {\bibinfo {title} {Wing--antenna interaction reduces odour fatigue in butterfly odour-tracking flight},}\ }\href@noop {} {\bibfield  {journal} {\bibinfo  {journal} {Journal of Fluid Mechanics}\ }\textbf {\bibinfo {volume} {998}},\ \bibinfo {pages} {A45} (\bibinfo {year} {2024})}\BibitemShut {NoStop}%
\bibitem [{\citenamefont {Peters}\ \emph {et~al.}(2019)\citenamefont {Peters}, \citenamefont {Peleg},\ and\ \citenamefont {Mahadevan}}]{peters2019collective}%
  \BibitemOpen
  \bibfield  {author} {\bibinfo {author} {\bibfnamefont {Jacob~M}\ \bibnamefont {Peters}}, \bibinfo {author} {\bibfnamefont {Orit}\ \bibnamefont {Peleg}}, \ and\ \bibinfo {author} {\bibfnamefont {L}~\bibnamefont {Mahadevan}},\ }\bibfield  {title} {\enquote {\bibinfo {title} {Collective ventilation in honeybee nests},}\ }\href@noop {} {\bibfield  {journal} {\bibinfo  {journal} {Journal of The Royal Society Interface}\ }\textbf {\bibinfo {volume} {16}},\ \bibinfo {pages} {20180561} (\bibinfo {year} {2019})}\BibitemShut {NoStop}%
\bibitem [{\citenamefont {Gilpin}\ \emph {et~al.}(2020)\citenamefont {Gilpin}, \citenamefont {Bull},\ and\ \citenamefont {Prakash}}]{gilpin2020multiscale}%
  \BibitemOpen
  \bibfield  {author} {\bibinfo {author} {\bibfnamefont {William}\ \bibnamefont {Gilpin}}, \bibinfo {author} {\bibfnamefont {Matthew~Storm}\ \bibnamefont {Bull}}, \ and\ \bibinfo {author} {\bibfnamefont {Manu}\ \bibnamefont {Prakash}},\ }\bibfield  {title} {\enquote {\bibinfo {title} {The multiscale physics of cilia and flagella},}\ }\href@noop {} {\bibfield  {journal} {\bibinfo  {journal} {Nature Reviews Physics}\ }\textbf {\bibinfo {volume} {2}},\ \bibinfo {pages} {74--88} (\bibinfo {year} {2020})}\BibitemShut {NoStop}%
\bibitem [{\citenamefont {Shelley}(2024)}]{shelley2024flows}%
  \BibitemOpen
  \bibfield  {author} {\bibinfo {author} {\bibfnamefont {Michael~J}\ \bibnamefont {Shelley}},\ }\bibfield  {title} {\enquote {\bibinfo {title} {Flows, self-organization, and transport in living cells},}\ }\href@noop {} {\bibfield  {journal} {\bibinfo  {journal} {Physical Review Fluids}\ }\textbf {\bibinfo {volume} {9}},\ \bibinfo {pages} {120501} (\bibinfo {year} {2024})}\BibitemShut {NoStop}%
\bibitem [{\citenamefont {den Toonder}\ \emph {et~al.}(2008)\citenamefont {den Toonder}, \citenamefont {Bos}, \citenamefont {Broer}, \citenamefont {Filippini}, \citenamefont {Gillies}, \citenamefont {de~Goede}, \citenamefont {Mol}, \citenamefont {Reijme}, \citenamefont {Talen}, \citenamefont {Wilderbeek} \emph {et~al.}}]{den2008artificial}%
  \BibitemOpen
  \bibfield  {author} {\bibinfo {author} {\bibfnamefont {Jaap}\ \bibnamefont {den Toonder}}, \bibinfo {author} {\bibfnamefont {Femke}\ \bibnamefont {Bos}}, \bibinfo {author} {\bibfnamefont {Dick}\ \bibnamefont {Broer}}, \bibinfo {author} {\bibfnamefont {Laura}\ \bibnamefont {Filippini}}, \bibinfo {author} {\bibfnamefont {Murray}\ \bibnamefont {Gillies}}, \bibinfo {author} {\bibfnamefont {Judith}\ \bibnamefont {de~Goede}}, \bibinfo {author} {\bibfnamefont {Titie}\ \bibnamefont {Mol}}, \bibinfo {author} {\bibfnamefont {Mireille}\ \bibnamefont {Reijme}}, \bibinfo {author} {\bibfnamefont {Wim}\ \bibnamefont {Talen}}, \bibinfo {author} {\bibfnamefont {Hans}\ \bibnamefont {Wilderbeek}},  \emph {et~al.},\ }\bibfield  {title} {\enquote {\bibinfo {title} {Artificial cilia for active micro-fluidic mixing},}\ }\href@noop {} {\bibfield  {journal} {\bibinfo  {journal} {Lab on a Chip}\ }\textbf {\bibinfo {volume} {8}},\ \bibinfo {pages} {533--541} (\bibinfo {year} {2008})}\BibitemShut {NoStop}%
\bibitem [{\citenamefont {Shields}\ \emph {et~al.}(2010)\citenamefont {Shields}, \citenamefont {Fiser}, \citenamefont {Evans}, \citenamefont {Falvo}, \citenamefont {Washburn},\ and\ \citenamefont {Superfine}}]{shields2010biomimetic}%
  \BibitemOpen
  \bibfield  {author} {\bibinfo {author} {\bibfnamefont {AR}~\bibnamefont {Shields}}, \bibinfo {author} {\bibfnamefont {BL}~\bibnamefont {Fiser}}, \bibinfo {author} {\bibfnamefont {EE}~\bibnamefont {Evans}}, \bibinfo {author} {\bibfnamefont {MR}~\bibnamefont {Falvo}}, \bibinfo {author} {\bibfnamefont {S}~\bibnamefont {Washburn}}, \ and\ \bibinfo {author} {\bibfnamefont {R}~\bibnamefont {Superfine}},\ }\bibfield  {title} {\enquote {\bibinfo {title} {Biomimetic cilia arrays generate simultaneous pumping and mixing regimes},}\ }\href@noop {} {\bibfield  {journal} {\bibinfo  {journal} {Proceedings of the National Academy of Sciences}\ }\textbf {\bibinfo {volume} {107}},\ \bibinfo {pages} {15670--15675} (\bibinfo {year} {2010})}\BibitemShut {NoStop}%
\bibitem [{\citenamefont {Wang}\ \emph {et~al.}(2023)\citenamefont {Wang}, \citenamefont {Sharma}, \citenamefont {Maldonado},\ and\ \citenamefont {Dong}}]{wang2023wirelessly}%
  \BibitemOpen
  \bibfield  {author} {\bibinfo {author} {\bibfnamefont {Yusheng}\ \bibnamefont {Wang}}, \bibinfo {author} {\bibfnamefont {Saksham}\ \bibnamefont {Sharma}}, \bibinfo {author} {\bibfnamefont {Fabien}\ \bibnamefont {Maldonado}}, \ and\ \bibinfo {author} {\bibfnamefont {Xiaoguang}\ \bibnamefont {Dong}},\ }\bibfield  {title} {\enquote {\bibinfo {title} {Wirelessly actuated ciliary airway stent for excessive mucus transportation},}\ }\href@noop {} {\bibfield  {journal} {\bibinfo  {journal} {Advanced materials technologies}\ }\textbf {\bibinfo {volume} {8}},\ \bibinfo {pages} {2301003} (\bibinfo {year} {2023})}\BibitemShut {NoStop}%
\bibitem [{\citenamefont {Zhao}\ \emph {et~al.}(2024)\citenamefont {Zhao}, \citenamefont {Zhao}, \citenamefont {Zhou},\ and\ \citenamefont {Chong}}]{zhao2024thermal}%
  \BibitemOpen
  \bibfield  {author} {\bibinfo {author} {\bibfnamefont {Hao-Bo}\ \bibnamefont {Zhao}}, \bibinfo {author} {\bibfnamefont {Chao-Ben}\ \bibnamefont {Zhao}}, \bibinfo {author} {\bibfnamefont {Quan}\ \bibnamefont {Zhou}}, \ and\ \bibinfo {author} {\bibfnamefont {Kai~Leong}\ \bibnamefont {Chong}},\ }\bibfield  {title} {\enquote {\bibinfo {title} {Thermal convection modulated by actively oscillating filament: The effect of filament rigidity},}\ }\href@noop {} {\bibfield  {journal} {\bibinfo  {journal} {International Journal of Heat and Mass Transfer}\ }\textbf {\bibinfo {volume} {228}},\ \bibinfo {pages} {125649} (\bibinfo {year} {2024})}\BibitemShut {NoStop}%
\bibitem [{\citenamefont {Nguyen}\ \emph {et~al.}(2021)\citenamefont {Nguyen}, \citenamefont {Liu}, \citenamefont {Raut}, \citenamefont {Bhattacharya}, \citenamefont {Sharma},\ and\ \citenamefont {Tran}}]{nguyen2021enhancement}%
  \BibitemOpen
  \bibfield  {author} {\bibinfo {author} {\bibfnamefont {Thien-Binh}\ \bibnamefont {Nguyen}}, \bibinfo {author} {\bibfnamefont {Dongdong}\ \bibnamefont {Liu}}, \bibinfo {author} {\bibfnamefont {Harshal}\ \bibnamefont {Raut}}, \bibinfo {author} {\bibfnamefont {Amitabh}\ \bibnamefont {Bhattacharya}}, \bibinfo {author} {\bibfnamefont {Atul}\ \bibnamefont {Sharma}}, \ and\ \bibinfo {author} {\bibfnamefont {Tuan}\ \bibnamefont {Tran}},\ }\bibfield  {title} {\enquote {\bibinfo {title} {Enhancement of convective heat transfer using magnetically flapping fin array},}\ }\href@noop {} {\bibfield  {journal} {\bibinfo  {journal} {International Communications in Heat and Mass Transfer}\ }\textbf {\bibinfo {volume} {129}},\ \bibinfo {pages} {105638} (\bibinfo {year} {2021})}\BibitemShut {NoStop}%
\bibitem [{\citenamefont {Xie}\ and\ \citenamefont {Zhang}(2020)}]{xie2020novel}%
  \BibitemOpen
  \bibfield  {author} {\bibinfo {author} {\bibfnamefont {Pengyong}\ \bibnamefont {Xie}}\ and\ \bibinfo {author} {\bibfnamefont {Xiaobing}\ \bibnamefont {Zhang}},\ }\bibfield  {title} {\enquote {\bibinfo {title} {A novel method of enhancing convective heat transfer by dynamic controlling rib},}\ }\href@noop {} {\bibfield  {journal} {\bibinfo  {journal} {International Communications in Heat and Mass Transfer}\ }\textbf {\bibinfo {volume} {119}},\ \bibinfo {pages} {104830} (\bibinfo {year} {2020})}\BibitemShut {NoStop}%
\bibitem [{\citenamefont {Jaffrin}(2012)}]{jaffrin2012hydrodynamic}%
  \BibitemOpen
  \bibfield  {author} {\bibinfo {author} {\bibfnamefont {Michel~Y}\ \bibnamefont {Jaffrin}},\ }\bibfield  {title} {\enquote {\bibinfo {title} {Hydrodynamic techniques to enhance membrane filtration},}\ }\href@noop {} {\bibfield  {journal} {\bibinfo  {journal} {Annual Review of Fluid Mechanics}\ }\textbf {\bibinfo {volume} {44}},\ \bibinfo {pages} {77--96} (\bibinfo {year} {2012})}\BibitemShut {NoStop}%
\bibitem [{\citenamefont {Turko}\ \emph {et~al.}(2020)\citenamefont {Turko}, \citenamefont {Cisternino},\ and\ \citenamefont {Wright}}]{turko2020calcified}%
  \BibitemOpen
  \bibfield  {author} {\bibinfo {author} {\bibfnamefont {Andy~J}\ \bibnamefont {Turko}}, \bibinfo {author} {\bibfnamefont {Bianca}\ \bibnamefont {Cisternino}}, \ and\ \bibinfo {author} {\bibfnamefont {Patricia~A}\ \bibnamefont {Wright}},\ }\bibfield  {title} {\enquote {\bibinfo {title} {Calcified gill filaments increase respiratory function in fishes},}\ }\href@noop {} {\bibfield  {journal} {\bibinfo  {journal} {Proceedings of the Royal Society B}\ }\textbf {\bibinfo {volume} {287}},\ \bibinfo {pages} {20192796} (\bibinfo {year} {2020})}\BibitemShut {NoStop}%
\bibitem [{\citenamefont {Strother}(2013)}]{strother2013hydrodynamic}%
  \BibitemOpen
  \bibfield  {author} {\bibinfo {author} {\bibfnamefont {James~A}\ \bibnamefont {Strother}},\ }\bibfield  {title} {\enquote {\bibinfo {title} {Hydrodynamic resistance and flow patterns in the gills of a tilapine fish},}\ }\href@noop {} {\bibfield  {journal} {\bibinfo  {journal} {Journal of Experimental Biology}\ }\textbf {\bibinfo {volume} {216}},\ \bibinfo {pages} {2595--2606} (\bibinfo {year} {2013})}\BibitemShut {NoStop}%
\bibitem [{\citenamefont {Smits}(2019)}]{smits2019undulatory}%
  \BibitemOpen
  \bibfield  {author} {\bibinfo {author} {\bibfnamefont {Alexander~J}\ \bibnamefont {Smits}},\ }\bibfield  {title} {\enquote {\bibinfo {title} {Undulatory and oscillatory swimming},}\ }\href@noop {} {\bibfield  {journal} {\bibinfo  {journal} {Journal of Fluid Mechanics}\ }\textbf {\bibinfo {volume} {874}},\ \bibinfo {pages} {P1} (\bibinfo {year} {2019})}\BibitemShut {NoStop}%
\bibitem [{\citenamefont {Dewey}\ \emph {et~al.}(2012)\citenamefont {Dewey}, \citenamefont {Carriou},\ and\ \citenamefont {Smits}}]{dewey2012relationship}%
  \BibitemOpen
  \bibfield  {author} {\bibinfo {author} {\bibfnamefont {Peter~A}\ \bibnamefont {Dewey}}, \bibinfo {author} {\bibfnamefont {Antoine}\ \bibnamefont {Carriou}}, \ and\ \bibinfo {author} {\bibfnamefont {Alexander~J}\ \bibnamefont {Smits}},\ }\bibfield  {title} {\enquote {\bibinfo {title} {On the relationship between efficiency and wake structure of a batoid-inspired oscillating fin},}\ }\href@noop {} {\bibfield  {journal} {\bibinfo  {journal} {Journal of fluid mechanics}\ }\textbf {\bibinfo {volume} {691}},\ \bibinfo {pages} {245--266} (\bibinfo {year} {2012})}\BibitemShut {NoStop}%
\bibitem [{\citenamefont {Moored}\ \emph {et~al.}(2012)\citenamefont {Moored}, \citenamefont {Dewey}, \citenamefont {Smits},\ and\ \citenamefont {Haj-Hariri}}]{moored2012hydrodynamic}%
  \BibitemOpen
  \bibfield  {author} {\bibinfo {author} {\bibfnamefont {Keith~W}\ \bibnamefont {Moored}}, \bibinfo {author} {\bibfnamefont {Peter~A}\ \bibnamefont {Dewey}}, \bibinfo {author} {\bibfnamefont {AJ}~\bibnamefont {Smits}}, \ and\ \bibinfo {author} {\bibfnamefont {H}~\bibnamefont {Haj-Hariri}},\ }\bibfield  {title} {\enquote {\bibinfo {title} {Hydrodynamic wake resonance as an underlying principle of efficient unsteady propulsion},}\ }\href@noop {} {\bibfield  {journal} {\bibinfo  {journal} {Journal of Fluid Mechanics}\ }\textbf {\bibinfo {volume} {708}},\ \bibinfo {pages} {329--348} (\bibinfo {year} {2012})}\BibitemShut {NoStop}%
\bibitem [{\citenamefont {Buchholz}\ and\ \citenamefont {Smits}(2006)}]{buchholz2006evolution}%
  \BibitemOpen
  \bibfield  {author} {\bibinfo {author} {\bibfnamefont {James~HJ}\ \bibnamefont {Buchholz}}\ and\ \bibinfo {author} {\bibfnamefont {Alexander~J}\ \bibnamefont {Smits}},\ }\bibfield  {title} {\enquote {\bibinfo {title} {On the evolution of the wake structure produced by a low-aspect-ratio pitching panel},}\ }\href@noop {} {\bibfield  {journal} {\bibinfo  {journal} {Journal of fluid mechanics}\ }\textbf {\bibinfo {volume} {546}},\ \bibinfo {pages} {433--443} (\bibinfo {year} {2006})}\BibitemShut {NoStop}%
\bibitem [{\citenamefont {Dong}\ \emph {et~al.}(2005)\citenamefont {Dong}, \citenamefont {Mittal}, \citenamefont {Bozkurttas},\ and\ \citenamefont {Najjar}}]{dong2005wake}%
  \BibitemOpen
  \bibfield  {author} {\bibinfo {author} {\bibfnamefont {Haibo}\ \bibnamefont {Dong}}, \bibinfo {author} {\bibfnamefont {Rajat}\ \bibnamefont {Mittal}}, \bibinfo {author} {\bibfnamefont {Meliha}\ \bibnamefont {Bozkurttas}}, \ and\ \bibinfo {author} {\bibfnamefont {Fady}\ \bibnamefont {Najjar}},\ }\bibfield  {title} {\enquote {\bibinfo {title} {Wake structure and performance of finite aspect-ratio flapping foils},}\ }in\ \href@noop {} {\emph {\bibinfo {booktitle} {43rd AIAA aerospace sciences meeting and exhibit}}}\ (\bibinfo {year} {2005})\ p.~\bibinfo {pages} {81}\BibitemShut {NoStop}%
\bibitem [{\citenamefont {Schnipper}\ \emph {et~al.}(2009)\citenamefont {Schnipper}, \citenamefont {Andersen},\ and\ \citenamefont {Bohr}}]{schnipper2009vortex}%
  \BibitemOpen
  \bibfield  {author} {\bibinfo {author} {\bibfnamefont {Teis}\ \bibnamefont {Schnipper}}, \bibinfo {author} {\bibfnamefont {Anders}\ \bibnamefont {Andersen}}, \ and\ \bibinfo {author} {\bibfnamefont {Tomas}\ \bibnamefont {Bohr}},\ }\bibfield  {title} {\enquote {\bibinfo {title} {Vortex wakes of a flapping foil},}\ }\href@noop {} {\bibfield  {journal} {\bibinfo  {journal} {Journal of Fluid Mechanics}\ }\textbf {\bibinfo {volume} {633}},\ \bibinfo {pages} {411--423} (\bibinfo {year} {2009})}\BibitemShut {NoStop}%
\bibitem [{\citenamefont {Van~Buren}\ \emph {et~al.}(2017)\citenamefont {Van~Buren}, \citenamefont {Floryan}, \citenamefont {Brunner}, \citenamefont {Senturk},\ and\ \citenamefont {Smits}}]{van2017impact}%
  \BibitemOpen
  \bibfield  {author} {\bibinfo {author} {\bibfnamefont {T}~\bibnamefont {Van~Buren}}, \bibinfo {author} {\bibfnamefont {D}~\bibnamefont {Floryan}}, \bibinfo {author} {\bibfnamefont {D}~\bibnamefont {Brunner}}, \bibinfo {author} {\bibfnamefont {U}~\bibnamefont {Senturk}}, \ and\ \bibinfo {author} {\bibfnamefont {AJ}~\bibnamefont {Smits}},\ }\bibfield  {title} {\enquote {\bibinfo {title} {Impact of trailing edge shape on the wake and propulsive performance of pitching panels},}\ }\href@noop {} {\bibfield  {journal} {\bibinfo  {journal} {Physical Review Fluids}\ }\textbf {\bibinfo {volume} {2}},\ \bibinfo {pages} {014702} (\bibinfo {year} {2017})}\BibitemShut {NoStop}%
\bibitem [{\citenamefont {Yeh}\ \emph {et~al.}(2017)\citenamefont {Yeh}, \citenamefont {Li},\ and\ \citenamefont {Alexeev}}]{yeh2017efficient}%
  \BibitemOpen
  \bibfield  {author} {\bibinfo {author} {\bibfnamefont {Peter~D}\ \bibnamefont {Yeh}}, \bibinfo {author} {\bibfnamefont {Yuanda}\ \bibnamefont {Li}}, \ and\ \bibinfo {author} {\bibfnamefont {Alexander}\ \bibnamefont {Alexeev}},\ }\bibfield  {title} {\enquote {\bibinfo {title} {Efficient swimming using flexible fins with tapered thickness},}\ }\href@noop {} {\bibfield  {journal} {\bibinfo  {journal} {Physical Review Fluids}\ }\textbf {\bibinfo {volume} {2}},\ \bibinfo {pages} {102101} (\bibinfo {year} {2017})}\BibitemShut {NoStop}%
\bibitem [{\citenamefont {Dean}\ and\ \citenamefont {Bhushan}(2010)}]{dean2010shark}%
  \BibitemOpen
  \bibfield  {author} {\bibinfo {author} {\bibfnamefont {Brian}\ \bibnamefont {Dean}}\ and\ \bibinfo {author} {\bibfnamefont {Bharat}\ \bibnamefont {Bhushan}},\ }\bibfield  {title} {\enquote {\bibinfo {title} {Shark-skin surfaces for fluid-drag reduction in turbulent flow: a review},}\ }\href@noop {} {\bibfield  {journal} {\bibinfo  {journal} {Philosophical Transactions of the Royal Society A: Mathematical, Physical and Engineering Sciences}\ }\textbf {\bibinfo {volume} {368}},\ \bibinfo {pages} {4775--4806} (\bibinfo {year} {2010})}\BibitemShut {NoStop}%
\bibitem [{\citenamefont {Bayazit}\ \emph {et~al.}(2014)\citenamefont {Bayazit}, \citenamefont {Sparrow},\ and\ \citenamefont {Joseph}}]{bayazit2014perforated}%
  \BibitemOpen
  \bibfield  {author} {\bibinfo {author} {\bibfnamefont {Yilmaz}\ \bibnamefont {Bayazit}}, \bibinfo {author} {\bibfnamefont {Eph~M}\ \bibnamefont {Sparrow}}, \ and\ \bibinfo {author} {\bibfnamefont {Daniel~D}\ \bibnamefont {Joseph}},\ }\bibfield  {title} {\enquote {\bibinfo {title} {Perforated plates for fluid management: Plate geometry effects and flow regimes},}\ }\href@noop {} {\bibfield  {journal} {\bibinfo  {journal} {International Journal of Thermal Sciences}\ }\textbf {\bibinfo {volume} {85}},\ \bibinfo {pages} {104--111} (\bibinfo {year} {2014})}\BibitemShut {NoStop}%
\bibitem [{\citenamefont {Marzin}\ \emph {et~al.}(2022)\citenamefont {Marzin}, \citenamefont {Le~Hay}, \citenamefont {de~Langre},\ and\ \citenamefont {Ramananarivo}}]{marzin2022flow}%
  \BibitemOpen
  \bibfield  {author} {\bibinfo {author} {\bibfnamefont {Tom}\ \bibnamefont {Marzin}}, \bibinfo {author} {\bibfnamefont {Kerian}\ \bibnamefont {Le~Hay}}, \bibinfo {author} {\bibfnamefont {Emmanuel}\ \bibnamefont {de~Langre}}, \ and\ \bibinfo {author} {\bibfnamefont {Sophie}\ \bibnamefont {Ramananarivo}},\ }\bibfield  {title} {\enquote {\bibinfo {title} {Flow-induced deformation of kirigami sheets},}\ }\href@noop {} {\bibfield  {journal} {\bibinfo  {journal} {Physical Review Fluids}\ }\textbf {\bibinfo {volume} {7}},\ \bibinfo {pages} {023906} (\bibinfo {year} {2022})}\BibitemShut {NoStop}%
\bibitem [{\citenamefont {Li}\ \emph {et~al.}(2021)\citenamefont {Li}, \citenamefont {Ran}, \citenamefont {Wang}, \citenamefont {Wang}, \citenamefont {Chen}, \citenamefont {Niu}, \citenamefont {Arratia},\ and\ \citenamefont {Yang}}]{li2021aerodynamics}%
  \BibitemOpen
  \bibfield  {author} {\bibinfo {author} {\bibfnamefont {Jing}\ \bibnamefont {Li}}, \bibinfo {author} {\bibfnamefont {Ranjiangshang}\ \bibnamefont {Ran}}, \bibinfo {author} {\bibfnamefont {Haihuan}\ \bibnamefont {Wang}}, \bibinfo {author} {\bibfnamefont {Yuchen}\ \bibnamefont {Wang}}, \bibinfo {author} {\bibfnamefont {You}\ \bibnamefont {Chen}}, \bibinfo {author} {\bibfnamefont {Shichao}\ \bibnamefont {Niu}}, \bibinfo {author} {\bibfnamefont {Paulo~E}\ \bibnamefont {Arratia}}, \ and\ \bibinfo {author} {\bibfnamefont {Shu}\ \bibnamefont {Yang}},\ }\bibfield  {title} {\enquote {\bibinfo {title} {Aerodynamics-assisted, efficient and scalable kirigami fog collectors},}\ }\href@noop {} {\bibfield  {journal} {\bibinfo  {journal} {Nature communications}\ }\textbf {\bibinfo {volume} {12}},\ \bibinfo {pages} {5484} (\bibinfo {year} {2021})}\BibitemShut {NoStop}%
\bibitem [{\citenamefont {Kakroo}\ and\ \citenamefont {Sadat}(2024)}]{kakroo2024high}%
  \BibitemOpen
  \bibfield  {author} {\bibinfo {author} {\bibfnamefont {Karan}\ \bibnamefont {Kakroo}}\ and\ \bibinfo {author} {\bibfnamefont {Hamid}\ \bibnamefont {Sadat}},\ }\bibfield  {title} {\enquote {\bibinfo {title} {High-fidelity fluid--structure interaction simulations of perforated elastic vortex generators},}\ }\href@noop {} {\bibfield  {journal} {\bibinfo  {journal} {Physics of Fluids}\ }\textbf {\bibinfo {volume} {36}} (\bibinfo {year} {2024})}\BibitemShut {NoStop}%
\bibitem [{\citenamefont {Guttag}\ \emph {et~al.}(2018)\citenamefont {Guttag}, \citenamefont {Karimi}, \citenamefont {Falc{\'o}n},\ and\ \citenamefont {Reis}}]{guttag2018aeroelastic}%
  \BibitemOpen
  \bibfield  {author} {\bibinfo {author} {\bibfnamefont {M}~\bibnamefont {Guttag}}, \bibinfo {author} {\bibfnamefont {Hussain~H}\ \bibnamefont {Karimi}}, \bibinfo {author} {\bibfnamefont {C}~\bibnamefont {Falc{\'o}n}}, \ and\ \bibinfo {author} {\bibfnamefont {Pedro~M}\ \bibnamefont {Reis}},\ }\bibfield  {title} {\enquote {\bibinfo {title} {Aeroelastic deformation of a perforated strip},}\ }\href@noop {} {\bibfield  {journal} {\bibinfo  {journal} {Physical Review Fluids}\ }\textbf {\bibinfo {volume} {3}},\ \bibinfo {pages} {014003} (\bibinfo {year} {2018})}\BibitemShut {NoStop}%
\bibitem [{\citenamefont {Jin}\ \emph {et~al.}(2020)\citenamefont {Jin}, \citenamefont {Kim}, \citenamefont {Cheng}, \citenamefont {Barry},\ and\ \citenamefont {Chamorro}}]{jin2020distinct}%
  \BibitemOpen
  \bibfield  {author} {\bibinfo {author} {\bibfnamefont {Yaqing}\ \bibnamefont {Jin}}, \bibinfo {author} {\bibfnamefont {Jin-Tae}\ \bibnamefont {Kim}}, \bibinfo {author} {\bibfnamefont {Shyuan}\ \bibnamefont {Cheng}}, \bibinfo {author} {\bibfnamefont {Oumar}\ \bibnamefont {Barry}}, \ and\ \bibinfo {author} {\bibfnamefont {Leonardo~P}\ \bibnamefont {Chamorro}},\ }\bibfield  {title} {\enquote {\bibinfo {title} {On the distinct drag, reconfiguration and wake of perforated structures},}\ }\href@noop {} {\bibfield  {journal} {\bibinfo  {journal} {Journal of Fluid Mechanics}\ }\textbf {\bibinfo {volume} {890}},\ \bibinfo {pages} {A1} (\bibinfo {year} {2020})}\BibitemShut {NoStop}%
\bibitem [{\citenamefont {Kasoju}\ and\ \citenamefont {Santhanakrishnan}(2021)}]{kasoju2021aerodynamic}%
  \BibitemOpen
  \bibfield  {author} {\bibinfo {author} {\bibfnamefont {Vishwa~T}\ \bibnamefont {Kasoju}}\ and\ \bibinfo {author} {\bibfnamefont {Arvind}\ \bibnamefont {Santhanakrishnan}},\ }\bibfield  {title} {\enquote {\bibinfo {title} {Aerodynamic interaction of bristled wing pairs in fling},}\ }\href@noop {} {\bibfield  {journal} {\bibinfo  {journal} {Physics of Fluids}\ }\textbf {\bibinfo {volume} {33}} (\bibinfo {year} {2021})}\BibitemShut {NoStop}%
\bibitem [{\citenamefont {Jones}\ \emph {et~al.}(2016)\citenamefont {Jones}, \citenamefont {Yun}, \citenamefont {Hedrick}, \citenamefont {Griffith},\ and\ \citenamefont {Miller}}]{jones2016bristles}%
  \BibitemOpen
  \bibfield  {author} {\bibinfo {author} {\bibfnamefont {Shannon~K}\ \bibnamefont {Jones}}, \bibinfo {author} {\bibfnamefont {Young~JJ}\ \bibnamefont {Yun}}, \bibinfo {author} {\bibfnamefont {Tyson~L}\ \bibnamefont {Hedrick}}, \bibinfo {author} {\bibfnamefont {Boyce~E}\ \bibnamefont {Griffith}}, \ and\ \bibinfo {author} {\bibfnamefont {Laura~A}\ \bibnamefont {Miller}},\ }\bibfield  {title} {\enquote {\bibinfo {title} {Bristles reduce the force required to ‘fling’wings apart in the smallest insects},}\ }\href@noop {} {\bibfield  {journal} {\bibinfo  {journal} {Journal of Experimental Biology}\ }\textbf {\bibinfo {volume} {219}},\ \bibinfo {pages} {3759--3772} (\bibinfo {year} {2016})}\BibitemShut {NoStop}%
\bibitem [{\citenamefont {Stamhuis}\ and\ \citenamefont {Thielicke}(2014)}]{stamhuis2014pivlab}%
  \BibitemOpen
  \bibfield  {author} {\bibinfo {author} {\bibfnamefont {Eize}\ \bibnamefont {Stamhuis}}\ and\ \bibinfo {author} {\bibfnamefont {William}\ \bibnamefont {Thielicke}},\ }\bibfield  {title} {\enquote {\bibinfo {title} {Pivlab--towards user-friendly, affordable and accurate digital particle image velocimetry in matlab},}\ }\href@noop {} {\bibfield  {journal} {\bibinfo  {journal} {Journal of open research software}\ }\textbf {\bibinfo {volume} {2}},\ \bibinfo {pages} {30} (\bibinfo {year} {2014})}\BibitemShut {NoStop}%
\bibitem [{\citenamefont {Shoele}\ and\ \citenamefont {Mittal}(2014)}]{shoele2014computational}%
  \BibitemOpen
  \bibfield  {author} {\bibinfo {author} {\bibfnamefont {Kourosh}\ \bibnamefont {Shoele}}\ and\ \bibinfo {author} {\bibfnamefont {Rajat}\ \bibnamefont {Mittal}},\ }\bibfield  {title} {\enquote {\bibinfo {title} {Computational study of flow-induced vibration of a reed in a channel and effect on convective heat transfer},}\ }\href@noop {} {\bibfield  {journal} {\bibinfo  {journal} {Physics of Fluids}\ }\textbf {\bibinfo {volume} {26}} (\bibinfo {year} {2014})}\BibitemShut {NoStop}%
\bibitem [{\citenamefont {Shampine}\ and\ \citenamefont {Reichelt}(1997)}]{shampine1997matlab}%
  \BibitemOpen
  \bibfield  {author} {\bibinfo {author} {\bibfnamefont {Lawrence~F}\ \bibnamefont {Shampine}}\ and\ \bibinfo {author} {\bibfnamefont {Mark~W}\ \bibnamefont {Reichelt}},\ }\bibfield  {title} {\enquote {\bibinfo {title} {The matlab ode suite},}\ }\href@noop {} {\bibfield  {journal} {\bibinfo  {journal} {SIAM journal on scientific computing}\ }\textbf {\bibinfo {volume} {18}},\ \bibinfo {pages} {1--22} (\bibinfo {year} {1997})}\BibitemShut {NoStop}%
\bibitem [{\citenamefont {Dormand}\ and\ \citenamefont {Prince}(1980)}]{dormand1980family}%
  \BibitemOpen
  \bibfield  {author} {\bibinfo {author} {\bibfnamefont {John~R}\ \bibnamefont {Dormand}}\ and\ \bibinfo {author} {\bibfnamefont {Peter~J}\ \bibnamefont {Prince}},\ }\bibfield  {title} {\enquote {\bibinfo {title} {A family of embedded runge-kutta formulae},}\ }\href@noop {} {\bibfield  {journal} {\bibinfo  {journal} {Journal of computational and applied mathematics}\ }\textbf {\bibinfo {volume} {6}},\ \bibinfo {pages} {19--26} (\bibinfo {year} {1980})}\BibitemShut {NoStop}%
\bibitem [{\citenamefont {Taneda}(1956)}]{taneda1956experimental}%
  \BibitemOpen
  \bibfield  {author} {\bibinfo {author} {\bibfnamefont {Sadatoshi}\ \bibnamefont {Taneda}},\ }\bibfield  {title} {\enquote {\bibinfo {title} {Experimental investigation of the wakes behind cylinders and plates at low reynolds numbers},}\ }\href@noop {} {\bibfield  {journal} {\bibinfo  {journal} {Journal of the Physical Society of Japan}\ }\textbf {\bibinfo {volume} {11}},\ \bibinfo {pages} {302--307} (\bibinfo {year} {1956})}\BibitemShut {NoStop}%
\bibitem [{\citenamefont {Park}\ \emph {et~al.}(2017)\citenamefont {Park}, \citenamefont {Chang}, \citenamefont {Kim},\ and\ \citenamefont {Sung}}]{park2017simulation}%
  \BibitemOpen
  \bibfield  {author} {\bibinfo {author} {\bibfnamefont {Sung~Goon}\ \bibnamefont {Park}}, \bibinfo {author} {\bibfnamefont {Cheong~Bong}\ \bibnamefont {Chang}}, \bibinfo {author} {\bibfnamefont {Boyoung}\ \bibnamefont {Kim}}, \ and\ \bibinfo {author} {\bibfnamefont {Hyung~Jin}\ \bibnamefont {Sung}},\ }\bibfield  {title} {\enquote {\bibinfo {title} {Simulation of fluid-flexible body interaction with heat transfer},}\ }\href@noop {} {\bibfield  {journal} {\bibinfo  {journal} {International Journal of Heat and Mass Transfer}\ }\textbf {\bibinfo {volume} {110}},\ \bibinfo {pages} {20--33} (\bibinfo {year} {2017})}\BibitemShut {NoStop}%
\bibitem [{\citenamefont {Dennis}\ \emph {et~al.}(1968)\citenamefont {Dennis}, \citenamefont {Hudson},\ and\ \citenamefont {Smith}}]{dennis1968steady}%
  \BibitemOpen
  \bibfield  {author} {\bibinfo {author} {\bibfnamefont {S~Co~R}\ \bibnamefont {Dennis}}, \bibinfo {author} {\bibfnamefont {JD}~\bibnamefont {Hudson}}, \ and\ \bibinfo {author} {\bibfnamefont {N}~\bibnamefont {Smith}},\ }\bibfield  {title} {\enquote {\bibinfo {title} {Steady laminar forced convection from a circular cylinder at low reynolds numbers},}\ }\href@noop {} {\bibfield  {journal} {\bibinfo  {journal} {The Physics of Fluids}\ }\textbf {\bibinfo {volume} {11}},\ \bibinfo {pages} {933--940} (\bibinfo {year} {1968})}\BibitemShut {NoStop}%
\bibitem [{\citenamefont {Hedrick}(2008)}]{hedrick2008software}%
  \BibitemOpen
  \bibfield  {author} {\bibinfo {author} {\bibfnamefont {Tyson~L}\ \bibnamefont {Hedrick}},\ }\bibfield  {title} {\enquote {\bibinfo {title} {Software techniques for two-and three-dimensional kinematic measurements of biological and biomimetic systems},}\ }\href@noop {} {\bibfield  {journal} {\bibinfo  {journal} {Bioinspiration \& biomimetics}\ }\textbf {\bibinfo {volume} {3}},\ \bibinfo {pages} {034001} (\bibinfo {year} {2008})}\BibitemShut {NoStop}%
\bibitem [{\citenamefont {Lekien}\ \emph {et~al.}(2005)\citenamefont {Lekien}, \citenamefont {Coulliette}, \citenamefont {Mariano}, \citenamefont {Ryan}, \citenamefont {Shay}, \citenamefont {Haller},\ and\ \citenamefont {Marsden}}]{lekien2005pollution}%
  \BibitemOpen
  \bibfield  {author} {\bibinfo {author} {\bibfnamefont {Francois}\ \bibnamefont {Lekien}}, \bibinfo {author} {\bibfnamefont {Chad}\ \bibnamefont {Coulliette}}, \bibinfo {author} {\bibfnamefont {Arthur~J}\ \bibnamefont {Mariano}}, \bibinfo {author} {\bibfnamefont {Edward~H}\ \bibnamefont {Ryan}}, \bibinfo {author} {\bibfnamefont {Lynn~K}\ \bibnamefont {Shay}}, \bibinfo {author} {\bibfnamefont {George}\ \bibnamefont {Haller}}, \ and\ \bibinfo {author} {\bibfnamefont {Jerry}\ \bibnamefont {Marsden}},\ }\bibfield  {title} {\enquote {\bibinfo {title} {Pollution release tied to invariant manifolds: A case study for the coast of florida},}\ }\href@noop {} {\bibfield  {journal} {\bibinfo  {journal} {Physica D: Nonlinear Phenomena}\ }\textbf {\bibinfo {volume} {210}},\ \bibinfo {pages} {1--20} (\bibinfo {year} {2005})}\BibitemShut {NoStop}%
\bibitem [{\citenamefont {Mathur}\ \emph {et~al.}(2007)\citenamefont {Mathur}, \citenamefont {Haller}, \citenamefont {Peacock}, \citenamefont {Ruppert-Felsot},\ and\ \citenamefont {Swinney}}]{mathur2007uncovering}%
  \BibitemOpen
  \bibfield  {author} {\bibinfo {author} {\bibfnamefont {Manikandan}\ \bibnamefont {Mathur}}, \bibinfo {author} {\bibfnamefont {George}\ \bibnamefont {Haller}}, \bibinfo {author} {\bibfnamefont {Thomas}\ \bibnamefont {Peacock}}, \bibinfo {author} {\bibfnamefont {Jori~E}\ \bibnamefont {Ruppert-Felsot}}, \ and\ \bibinfo {author} {\bibfnamefont {Harry~L}\ \bibnamefont {Swinney}},\ }\bibfield  {title} {\enquote {\bibinfo {title} {Uncovering the lagrangian skeleton of turbulence},}\ }\href@noop {} {\bibfield  {journal} {\bibinfo  {journal} {Physical Review Letters}\ }\textbf {\bibinfo {volume} {98}},\ \bibinfo {pages} {144502} (\bibinfo {year} {2007})}\BibitemShut {NoStop}%
\bibitem [{\citenamefont {Peng}\ and\ \citenamefont {Dabiri}(2009)}]{peng2009transport}%
  \BibitemOpen
  \bibfield  {author} {\bibinfo {author} {\bibfnamefont {J}~\bibnamefont {Peng}}\ and\ \bibinfo {author} {\bibfnamefont {JO}~\bibnamefont {Dabiri}},\ }\bibfield  {title} {\enquote {\bibinfo {title} {Transport of inertial particles by lagrangian coherent structures: application to predator--prey interaction in jellyfish feeding},}\ }\href@noop {} {\bibfield  {journal} {\bibinfo  {journal} {Journal of Fluid Mechanics}\ }\textbf {\bibinfo {volume} {623}},\ \bibinfo {pages} {75--84} (\bibinfo {year} {2009})}\BibitemShut {NoStop}%
\bibitem [{\citenamefont {Wu}\ \emph {et~al.}(2024)\citenamefont {Wu}, \citenamefont {Basu}, \citenamefont {Kim}, \citenamefont {Sorrells}, \citenamefont {Beron-Vera},\ and\ \citenamefont {Jung}}]{wu2024coherent}%
  \BibitemOpen
  \bibfield  {author} {\bibinfo {author} {\bibfnamefont {Zixuan}\ \bibnamefont {Wu}}, \bibinfo {author} {\bibfnamefont {Saikat}\ \bibnamefont {Basu}}, \bibinfo {author} {\bibfnamefont {Seungho}\ \bibnamefont {Kim}}, \bibinfo {author} {\bibfnamefont {Mark}\ \bibnamefont {Sorrells}}, \bibinfo {author} {\bibfnamefont {Francisco~J}\ \bibnamefont {Beron-Vera}}, \ and\ \bibinfo {author} {\bibfnamefont {Sunghwan}\ \bibnamefont {Jung}},\ }\bibfield  {title} {\enquote {\bibinfo {title} {Coherent spore dispersion via drop-leaf interaction},}\ }\href@noop {} {\bibfield  {journal} {\bibinfo  {journal} {Science advances}\ }\textbf {\bibinfo {volume} {10}},\ \bibinfo {pages} {eadj8092} (\bibinfo {year} {2024})}\BibitemShut {NoStop}%
\bibitem [{\citenamefont {Haller}(2015)}]{haller2015lagrangian}%
  \BibitemOpen
  \bibfield  {author} {\bibinfo {author} {\bibfnamefont {George}\ \bibnamefont {Haller}},\ }\bibfield  {title} {\enquote {\bibinfo {title} {Lagrangian coherent structures},}\ }\href@noop {} {\bibfield  {journal} {\bibinfo  {journal} {Annual review of fluid mechanics}\ }\textbf {\bibinfo {volume} {47}},\ \bibinfo {pages} {137--162} (\bibinfo {year} {2015})}\BibitemShut {NoStop}%
\bibitem [{\citenamefont {Shadden}\ \emph {et~al.}(2005)\citenamefont {Shadden}, \citenamefont {Lekien},\ and\ \citenamefont {Marsden}}]{shadden2005definition}%
  \BibitemOpen
  \bibfield  {author} {\bibinfo {author} {\bibfnamefont {Shawn~C}\ \bibnamefont {Shadden}}, \bibinfo {author} {\bibfnamefont {Francois}\ \bibnamefont {Lekien}}, \ and\ \bibinfo {author} {\bibfnamefont {Jerrold~E}\ \bibnamefont {Marsden}},\ }\bibfield  {title} {\enquote {\bibinfo {title} {Definition and properties of lagrangian coherent structures from finite-time lyapunov exponents in two-dimensional aperiodic flows},}\ }\href@noop {} {\bibfield  {journal} {\bibinfo  {journal} {Physica D: Nonlinear Phenomena}\ }\textbf {\bibinfo {volume} {212}},\ \bibinfo {pages} {271--304} (\bibinfo {year} {2005})}\BibitemShut {NoStop}%
\bibitem [{\citenamefont {Haller}\ and\ \citenamefont {Yuan}(2000)}]{haller2000lagrangian}%
  \BibitemOpen
  \bibfield  {author} {\bibinfo {author} {\bibfnamefont {George}\ \bibnamefont {Haller}}\ and\ \bibinfo {author} {\bibfnamefont {Guocheng}\ \bibnamefont {Yuan}},\ }\bibfield  {title} {\enquote {\bibinfo {title} {Lagrangian coherent structures and mixing in two-dimensional turbulence},}\ }\href@noop {} {\bibfield  {journal} {\bibinfo  {journal} {Physica D: Nonlinear Phenomena}\ }\textbf {\bibinfo {volume} {147}},\ \bibinfo {pages} {352--370} (\bibinfo {year} {2000})}\BibitemShut {NoStop}%
\bibitem [{\citenamefont {Brunton}\ and\ \citenamefont {Rowley}(2010)}]{brunton2010fast}%
  \BibitemOpen
  \bibfield  {author} {\bibinfo {author} {\bibfnamefont {Steven~L}\ \bibnamefont {Brunton}}\ and\ \bibinfo {author} {\bibfnamefont {Clarence~W}\ \bibnamefont {Rowley}},\ }\bibfield  {title} {\enquote {\bibinfo {title} {Fast computation of finite-time lyapunov exponent fields for unsteady flows},}\ }\href@noop {} {\bibfield  {journal} {\bibinfo  {journal} {Chaos: An Interdisciplinary Journal of Nonlinear Science}\ }\textbf {\bibinfo {volume} {20}} (\bibinfo {year} {2010})}\BibitemShut {NoStop}%
\bibitem [{\citenamefont {Shadden}\ \emph {et~al.}(2006)\citenamefont {Shadden}, \citenamefont {Dabiri},\ and\ \citenamefont {Marsden}}]{shadden2006lagrangian}%
  \BibitemOpen
  \bibfield  {author} {\bibinfo {author} {\bibfnamefont {Shawn~C}\ \bibnamefont {Shadden}}, \bibinfo {author} {\bibfnamefont {John~O}\ \bibnamefont {Dabiri}}, \ and\ \bibinfo {author} {\bibfnamefont {Jerrold~E}\ \bibnamefont {Marsden}},\ }\bibfield  {title} {\enquote {\bibinfo {title} {Lagrangian analysis of fluid transport in empirical vortex ring flows},}\ }\href@noop {} {\bibfield  {journal} {\bibinfo  {journal} {Physics of fluids}\ }\textbf {\bibinfo {volume} {18}} (\bibinfo {year} {2006})}\BibitemShut {NoStop}%
\bibitem [{\citenamefont {Voth}\ \emph {et~al.}(2002)\citenamefont {Voth}, \citenamefont {Haller},\ and\ \citenamefont {Gollub}}]{voth2002experimental}%
  \BibitemOpen
  \bibfield  {author} {\bibinfo {author} {\bibfnamefont {Greg~A}\ \bibnamefont {Voth}}, \bibinfo {author} {\bibfnamefont {G}~\bibnamefont {Haller}}, \ and\ \bibinfo {author} {\bibfnamefont {Jerry~P}\ \bibnamefont {Gollub}},\ }\bibfield  {title} {\enquote {\bibinfo {title} {Experimental measurements of stretching fields in fluid mixing},}\ }\href@noop {} {\bibfield  {journal} {\bibinfo  {journal} {Physical review letters}\ }\textbf {\bibinfo {volume} {88}},\ \bibinfo {pages} {254501} (\bibinfo {year} {2002})}\BibitemShut {NoStop}%
\bibitem [{\citenamefont {Liu}\ \emph {et~al.}(1994)\citenamefont {Liu}, \citenamefont {Muzzio},\ and\ \citenamefont {Peskin}}]{liu1994quantification}%
  \BibitemOpen
  \bibfield  {author} {\bibinfo {author} {\bibfnamefont {M}~\bibnamefont {Liu}}, \bibinfo {author} {\bibfnamefont {FJ}~\bibnamefont {Muzzio}}, \ and\ \bibinfo {author} {\bibfnamefont {RL}~\bibnamefont {Peskin}},\ }\bibfield  {title} {\enquote {\bibinfo {title} {Quantification of mixing in aperiodic chaotic flows},}\ }\href@noop {} {\bibfield  {journal} {\bibinfo  {journal} {Chaos, Solitons \& Fractals}\ }\textbf {\bibinfo {volume} {4}},\ \bibinfo {pages} {869--893} (\bibinfo {year} {1994})}\BibitemShut {NoStop}%
\bibitem [{\citenamefont {Badas}\ \emph {et~al.}(2017)\citenamefont {Badas}, \citenamefont {Domenichini},\ and\ \citenamefont {Querzoli}}]{badas2017quantification}%
  \BibitemOpen
  \bibfield  {author} {\bibinfo {author} {\bibfnamefont {MARIA~GRAZIA}\ \bibnamefont {Badas}}, \bibinfo {author} {\bibfnamefont {Federico}\ \bibnamefont {Domenichini}}, \ and\ \bibinfo {author} {\bibfnamefont {Giorgio}\ \bibnamefont {Querzoli}},\ }\bibfield  {title} {\enquote {\bibinfo {title} {Quantification of the blood mixing in the left ventricle using finite time lyapunov exponents},}\ }\href@noop {} {\bibfield  {journal} {\bibinfo  {journal} {Meccanica}\ }\textbf {\bibinfo {volume} {52}},\ \bibinfo {pages} {529--544} (\bibinfo {year} {2017})}\BibitemShut {NoStop}%
\bibitem [{\citenamefont {Beron-Vera}\ \emph {et~al.}(2010)\citenamefont {Beron-Vera}, \citenamefont {Olascoaga},\ and\ \citenamefont {Goni}}]{beron2010surface}%
  \BibitemOpen
  \bibfield  {author} {\bibinfo {author} {\bibfnamefont {Francisco~J}\ \bibnamefont {Beron-Vera}}, \bibinfo {author} {\bibfnamefont {Mar{\'\i}a~J}\ \bibnamefont {Olascoaga}}, \ and\ \bibinfo {author} {\bibfnamefont {Gustavo~J}\ \bibnamefont {Goni}},\ }\bibfield  {title} {\enquote {\bibinfo {title} {Surface ocean mixing inferred from different multisatellite altimetry measurements},}\ }\href@noop {} {\bibfield  {journal} {\bibinfo  {journal} {Journal of physical oceanography}\ }\textbf {\bibinfo {volume} {40}},\ \bibinfo {pages} {2466--2480} (\bibinfo {year} {2010})}\BibitemShut {NoStop}%
\bibitem [{\citenamefont {Lambert}\ and\ \citenamefont {Rangel}(2010)}]{lambert2010role}%
  \BibitemOpen
  \bibfield  {author} {\bibinfo {author} {\bibfnamefont {Ruth~A}\ \bibnamefont {Lambert}}\ and\ \bibinfo {author} {\bibfnamefont {Roger~H}\ \bibnamefont {Rangel}},\ }\bibfield  {title} {\enquote {\bibinfo {title} {The role of elastic flap deformation on fluid mixing in a microchannel},}\ }\href@noop {} {\bibfield  {journal} {\bibinfo  {journal} {Physics of Fluids}\ }\textbf {\bibinfo {volume} {22}} (\bibinfo {year} {2010})}\BibitemShut {NoStop}%
\bibitem [{\citenamefont {Saini}\ \emph {et~al.}(2023)\citenamefont {Saini}, \citenamefont {Dhar},\ and\ \citenamefont {Powar}}]{saini2023performance}%
  \BibitemOpen
  \bibfield  {author} {\bibinfo {author} {\bibfnamefont {Prashant}\ \bibnamefont {Saini}}, \bibinfo {author} {\bibfnamefont {Atul}\ \bibnamefont {Dhar}}, \ and\ \bibinfo {author} {\bibfnamefont {Satvasheel}\ \bibnamefont {Powar}},\ }\bibfield  {title} {\enquote {\bibinfo {title} {Performance enhancement of fin and tube heat exchanger employing curved trapezoidal winglet vortex generator with circular punched holes},}\ }\href@noop {} {\bibfield  {journal} {\bibinfo  {journal} {International Journal of Heat and Mass Transfer}\ }\textbf {\bibinfo {volume} {209}},\ \bibinfo {pages} {124142} (\bibinfo {year} {2023})}\BibitemShut {NoStop}%
\bibitem [{\citenamefont {Quinn}\ \emph {et~al.}(2014)\citenamefont {Quinn}, \citenamefont {Lauder},\ and\ \citenamefont {Smits}}]{quinn2014scaling}%
  \BibitemOpen
  \bibfield  {author} {\bibinfo {author} {\bibfnamefont {Daniel~B}\ \bibnamefont {Quinn}}, \bibinfo {author} {\bibfnamefont {George~V}\ \bibnamefont {Lauder}}, \ and\ \bibinfo {author} {\bibfnamefont {Alexander~J}\ \bibnamefont {Smits}},\ }\bibfield  {title} {\enquote {\bibinfo {title} {Scaling the propulsive performance of heaving flexible panels},}\ }\href@noop {} {\bibfield  {journal} {\bibinfo  {journal} {Journal of fluid mechanics}\ }\textbf {\bibinfo {volume} {738}},\ \bibinfo {pages} {250--267} (\bibinfo {year} {2014})}\BibitemShut {NoStop}%
\bibitem [{\citenamefont {Liu}\ \emph {et~al.}(2022)\citenamefont {Liu}, \citenamefont {Zhong}, \citenamefont {Han}, \citenamefont {Moored},\ and\ \citenamefont {Quinn}}]{liu2022fine}%
  \BibitemOpen
  \bibfield  {author} {\bibinfo {author} {\bibfnamefont {Leo}\ \bibnamefont {Liu}}, \bibinfo {author} {\bibfnamefont {Qiang}\ \bibnamefont {Zhong}}, \bibinfo {author} {\bibfnamefont {Tianjun}\ \bibnamefont {Han}}, \bibinfo {author} {\bibfnamefont {Keith~W}\ \bibnamefont {Moored}}, \ and\ \bibinfo {author} {\bibfnamefont {Daniel~B}\ \bibnamefont {Quinn}},\ }\bibfield  {title} {\enquote {\bibinfo {title} {Fine-tuning near-boundary swimming equilibria using asymmetric kinematics},}\ }\href@noop {} {\bibfield  {journal} {\bibinfo  {journal} {Bioinspiration \& Biomimetics}\ }\textbf {\bibinfo {volume} {18}},\ \bibinfo {pages} {016011} (\bibinfo {year} {2022})}\BibitemShut {NoStop}%
\bibitem [{\citenamefont {Shang}\ \emph {et~al.}(2018)\citenamefont {Shang}, \citenamefont {Stone},\ and\ \citenamefont {Smits}}]{shang2018flow}%
  \BibitemOpen
  \bibfield  {author} {\bibinfo {author} {\bibfnamefont {Jessica~K}\ \bibnamefont {Shang}}, \bibinfo {author} {\bibfnamefont {Howard~A}\ \bibnamefont {Stone}}, \ and\ \bibinfo {author} {\bibfnamefont {AJ}~\bibnamefont {Smits}},\ }\bibfield  {title} {\enquote {\bibinfo {title} {Flow past finite cylinders of constant curvature},}\ }\href@noop {} {\bibfield  {journal} {\bibinfo  {journal} {Journal of Fluid Mechanics}\ }\textbf {\bibinfo {volume} {837}},\ \bibinfo {pages} {896--915} (\bibinfo {year} {2018})}\BibitemShut {NoStop}%
\bibitem [{\citenamefont {Park}\ \emph {et~al.}(2014)\citenamefont {Park}, \citenamefont {Kim},\ and\ \citenamefont {Kim}}]{park2014optimal}%
  \BibitemOpen
  \bibfield  {author} {\bibinfo {author} {\bibfnamefont {Keunhwan}\ \bibnamefont {Park}}, \bibinfo {author} {\bibfnamefont {Wonjung}\ \bibnamefont {Kim}}, \ and\ \bibinfo {author} {\bibfnamefont {Ho-Young}\ \bibnamefont {Kim}},\ }\bibfield  {title} {\enquote {\bibinfo {title} {Optimal lamellar arrangement in fish gills},}\ }\href@noop {} {\bibfield  {journal} {\bibinfo  {journal} {Proceedings of the National Academy of Sciences}\ }\textbf {\bibinfo {volume} {111}},\ \bibinfo {pages} {8067--8070} (\bibinfo {year} {2014})}\BibitemShut {NoStop}%
\end{thebibliography}%
\end{document}